\documentclass[letterpaper]{article} 
\usepackage{aaai23}  
\usepackage{times}  
\usepackage{helvet}  
\usepackage{courier}  
\usepackage[hyphens]{url}  
\usepackage{graphicx} 
\usepackage{natbib}  
\usepackage{caption} 
\DeclareCaptionStyle{ruled}{labelfont=normalfont,labelsep=colon,strut=off} 

\usepackage{amsmath}
\usepackage{amsthm}
\usepackage{amssymb}
\usepackage{thm-restate}

\newtheorem{theorem}{Theorem}
\newtheorem{definition}{Definition}
\newtheorem{proposition}{Proposition}
\newtheorem{lemma}{Lemma}
\newtheorem{corollary}{Corollary}

\usepackage{algorithm}
\usepackage{algorithmic}
\usepackage{lipsum}
\usepackage{multirow}
\usepackage[table]{xcolor}
\usepackage{booktabs} 
\usepackage[capitalise]{cleveref}
\usepackage[all]{foreign}
\usepackage[inline]{enumitem}

\usepackage{packages/commands} 
\usepackage{packages/logic}    
\usepackage{packages/symbols}  

\newcommand{\nb}[1]{}

\allowdisplaybreaks

\newif\ifblind
\blindfalse

\newif\ifappendix
\appendixtrue

\IfFileExists{config.cfg}{\input{config.cfg}}

\title{Complexity of Safety and coSafety Fragments of Linear Temporal Logic}
\ifblind
  \author{Submission \#8719}
\else
  \author{
      Alessandro Artale\textsuperscript{\rm 1},
      Luca Geatti\textsuperscript{\rm 2}\thanks{Corresponding author.},
      Nicola Gigante\textsuperscript{\rm 1},
      Andrea Mazzullo\textsuperscript{\rm 1}\footnotemark[1],
      Angelo Montanari\textsuperscript{\rm 2}
  }
  \affiliations{
      \textsuperscript{\rm 1}Free University of Bozen-Bolzano\\
      \textsuperscript{\rm 2} University of Udine
  
      artale@inf.unibz.it,
      luca.geatti@uniud.it,
      gigante@inf.unibz.it,
      mazzullo@inf.unibz.it,
      angelo.montanari@uniud.it
  }
\fi

\begin{document}

\maketitle

\begin{abstract}
  Linear Temporal Logic (\LTL) is the de-facto standard temporal logic for
  system specification, whose foundational properties have been studied for over
  five decades. Safety and cosafety properties define 
  notable fragments of \LTL, where 
  a prefix of a trace suffices to establish whether a formula is true or not
  over that trace.
  In this paper, we study the complexity of the problems of satisfiability, 
  validity, and realizability over infinite and finite traces for the 
  safety and cosafety fragments of \LTL.
  As for satisfiability and validity over infinite traces, we prove that the
  majority of the fragments have the same complexity as full \LTL, that is,
  they are \PSPACE-complete.
  The picture is radically different for realizability: we find 
  fragments 
  with the same expressive power 
  whose complexity 
  varies from
  \EXPTIME[2]-complete (as full \LTL) to \EXPTIME-complete.
  Notably, for all cosafety fragments, the complexity of the
  three problems 
  does not change 
  passing from infinite to finite traces, while
  for all safety fragments the complexity of satisfiability
  (resp.,  realizability) over finite traces drops to \NP-complete (resp.,
  $\Pi^P_2$-complete).
\end{abstract}



\section{Introduction}
\label{sec:intro}

Linear Temporal Logic (\LTL) is arguably
the most
renowned
temporal logic,
with applications in a variety of branches of computer
science~\cite{pnueli1977temporal,vardi1986automata,vardi1994reasoning}.
\LTL is usually interpreted
over
infinite state sequences (or traces);
recently, the finite-trace semantics has received
attention as well, especially in \emph{artificial intelligence}~\cite{DeGiacomoV13,DeGiacomoV15,DegEtAl,FiondaG18,AMO:IJCAI19}.

The satisfiability (resp., validity) problem of \LTL consists of deciding
whether, given an \LTL formula, it is satisfied by at least one state
sequence (resp., by all state sequences). It is known that satisfiability
and validity of \LTL, interpreted  over both infinite and finite traces, are
\PSPACE-complete~\cite{sistla1985complexity,DeGiacomoV13}.
Realizability~\cite{pnueli1989synthesis} is 
more complex 
than satisfiability:
it asks, for a given formula over a set of variables partitioned into controllable and uncontrollable ones, whether there exists a strategy such that, for any value of the uncontrollable variables, chooses the value of the controllable ones in such a way to satisfy the formula.
\LTL realizability is \EXPTIME[2]-complete, on both
infinite~\cite{rosner1992modular} and finite~\cite{DeGiacomoV15} traces.

Despite the complexity of these problems, several \LTL tools have been
developed, including model checkers and translators to automata.  
However, 
some applications (such as in \emph{runtime verification})
do not always require
the full expressivity of \LTL,
and would rather benefit instead from computational efficiency.
Several fragments considered in the literature address these aspects.
Two
notable ones are the \emph{safety} and \emph{cosafety}
fragments~\cite{sistla1994safety}: they are a subclass of $\omega$-regular
languages where a finite prefix suffices to establish the 
membership of an infinite word to a language,
thus allowing one to reason over finite traces.
This
is very helpful in practice, \eg it allows one to avoid Safra's
determinization algorithm~\cite{safra1988complexity}
in favor of the classical subset construction.
However, 
despite their usefulness, a systematic complexity analysis of reasoning in these fragments, over both infinite and finite traces, is missing.

In this paper, we study the complexity of the  satisfiability,
validity, and realizability problems for 
safety and cosafety \LTL fragments, 
over both infinite and finite traces. We focus on three
cosafety fragments and the dual safety ones, some of which are
expressively complete with respect to the set of \LTL-definable (co)safety
properties.

We first prove that the complexity of the satisfiability and validity problems for the majority of the considered fragments (both safety and cosafety) is the same as full \LTL, that is, \PSPACE-complete.

As for cosafety fragments, we prove a general theorem that allows us to
transfer all complexity results for satisfiability, validity, and
realizability from infinite to finite trace semantics.  On the contrary, the difference in complexity when passing from
infinite to finite traces is not negligible for safety fragments. 
We prove a small (bounded) model property for all safety regular languages of finite words, which states that if a language is not empty, then it contains a word of length $1$.
By exploiting this result, we show that the complexity of satisfiability (resp.,
realizability) of all safety fragments drops to \NP-complete (resp.,
$\Pi^P_2$-complete) when considering finite traces.

Finally, we show that some of the fragments, although being expressively
equivalent, have different complexities for realizability. In particular,
for fragments that use past modalities (or are devoid of \emph{until} and \emph{release} modalities), the complexity turns out to be
\EXPTIME-complete, in contrast to the \EXPTIME[2]-completeness of the other
fragments. 

The paper is organized as follows. In Sec.~2, we provide the necessary
background. Sec.~3 contains two general theorems that we will use for
establishing the complexities of the fragments under finite trace
semantics. Sec.~4 and Sec.~5 study the complexity of the fragments, over both
infinite and finite traces, for the satisfiability/validity and
realizability problems, respectively. In Sec.~6 we discuss the results, while in Sec.~7 we point out future research directions.

\ifappendix
The proof of all Lemmata and Theorems, if not present in the body of the
paper, are given in the Appendix.
We also include an Erratum that fixes some problems, pointed out by N.~Arteche, in the proofs of~\cref{sat:cosafetyltl} and~\cref{real:cosafetyltl} of the original version of this paper.
\else
The proof of all Lemmata and Theorems can be found in~\cite{artaleetal2022arxiv}.
\fi


\section{Preliminaries}
\label{sec:prel}

In this section, we provide the necessary background.

\paragraph{Linear Temporal Logic.}
Given a set $\Sigma$ of proposition letters, an \emph{\LTLP formula} $\phi$
is generated as follows: 
\begin{align*}
\phi
\coloneqq 
p
& \mid \neg p \mid \phi \lor \phi \mid \phi \land \phi & \text{Boolean connectives} \\
& \mid \ltl{X} \phi \mid \ltl{wX} \phi \mid \phi \ltl{U} \phi \mid \phi \ltl{R} \phi & \text{future modalities} \\
& \mid \ltl{Y} \phi \mid \ltl{wY} \phi \mid \phi \ltl{S} \phi \mid \phi \ltl{T} \phi & \text{past modalities}
\end{align*}
where $p \in \Sigma$.  We use the standard
shortcuts for $\top \coloneqq p \lor
\lnot p$, $\bot \coloneqq p \land \lnot p$ (for some $p \in
\Sigma$) and other temporal operators: $\ltl{F\phi \coloneqq \top U \phi}$,
$\ltl{G\phi \coloneqq \false R \phi}$, $\ltl{P\phi \coloneqq \top S \phi}$, and $\ltl{H\phi
\coloneqq \false T \phi}$. 
Note that, \emph{w.l.o.g.}, our definition of \LTLP considers
formulas already in Negation Normal Form (NNF), that is, negations are
applied only to proposition letters. 

A \emph{pure future} (resp., \emph{past}) \emph{formula} is an \LTLP
formula without occurrences of past (resp., future) modalities. We denote by
\LTL (resp., \LTLFP) the set of pure future (resp.,  past) formulas.
Given a set $S \subseteq \set{\ltl{X,wX,U,R,Y,wY,S,T}}$ of temporal operators
and $\mathbb{L} \in \{ \LTLP, \LTL, \LTLFP \}$,
we denote by
$\mathbb{L}[S]$
the set of formulas $\phi$ of
$\mathbb{L}$
restricted\nb{A: changed} to operators in $S$.
In the following, we denote by \safetyltl (resp., \cosafetyltl) the fragment $\LTL[\ltl{wX,R}]$ (resp., $\LTL[\ltl{X,U}]$), also known as the \emph{syntactic (co-)safety fragment of
\LTL}~\cite{sistla1985characterization,ChangMP92,ZhuTLPV17}.
Finally, we denote by $\Galpha$ (resp., $\Falpha$) the
set of $\LTLP$ formulas of the form $\ltl{G\alpha}$ (resp.,
$\ltl{F\alpha}$), with $\alpha \in \LTLFP$.

%

Let $\sigma \in (2^{\Sigma})^{+} \cup (2^{\Sigma})^{\omega}$ be
a \emph{state sequence} (or \emph{trace}, or \emph{word}).
We define the \emph{length} of $\sigma$ as $|\sigma| = n$, if $\sigma = \langle \sigma_{0}, \ldots, \sigma_{n-1} \rangle \in (2^{\Sigma})^{+}$; $| \sigma | = \omega$, if $\sigma \in (2^{\Sigma})^{\omega}$.
The \emph{satisfaction} of an
\LTLP formula $\phi$ by $\sigma$ at time $0 \leq i < |\sigma|$, denoted by $\sigma,
i \models \phi$, is defined as follows
(we omit Booleans):

\begin{itemize}
  \item $\sigma,i \models p$ iff $p\in\state_i$;
  \item $\sigma,i \models \ltl{X\phi}$     iff 
          $i+1<|\sigma|$ and  $\sigma,i+1\models \phi$;
  \item $\sigma,i \models \ltl{wX\phi}$     iff 
          either $i+1=|\sigma|$ or $\sigma,i+1\models \phi$;
  \item $\sigma,i \models \ltl{Y\phi}$    iff
          $i > 0$ and $\sigma,i-1\models \phi$;
  \item $\sigma,i \models \ltl{wY\phi}$    iff
          either $i = 0$ or $\sigma,i-1\models \phi$;
  \item $\sigma,i \models \ltl{\phi_1 U \phi_2}$  iff
          there exists $i\le j<|\sigma|$ such that $\sigma,j\models\phi_2$,
          and $\sigma,k\models\phi_1$ for all $k$, with $i \le k < j$;
  \item $\sigma,i \models \ltl{\phi_1 S \phi_2}$    iff
          there exists $j\le i$ such that $\sigma,j\models\phi_2$,
          and $\sigma,k\models\phi_1$ for all $k$, with $j < k \le i$;
  \item $\sigma,i \models \ltl{\phi_1 R \phi_2}$  iff either
    $\sigma,j\models\phi_2$ for all $i\le j < |\sigma|$, or there exists
    $i \leq k < | \sigma |$ such that $\sigma,k\models\phi_1$ and $\sigma,j\models\phi_2$
    for all $i\le j \le k$;
  \item $\sigma,i \models \ltl{\phi_1 T \phi_2}$  iff either
    $\sigma,j\models\phi_2$ for all $0\le j \leq i$, or there exists $k \le
    i$ such that $\sigma,k\models\phi_1$ and $\sigma,j\models\phi_2$ for
    all $i\ge j \ge k$.
\end{itemize}

We say that $\sigma$ is a \emph{model} of $\phi$ (written as $\sigma
\models \phi$) iff $\sigma,0 \models \phi$.
The \emph{language of infinite} (resp., \emph{finite}) \emph{traces} of
$\phi$, denoted by $\lang(\phi)$, is the set of traces
$\sigma\in(2^\Sigma)^\omega$ (resp., $\sigma\in(2^\Sigma)^+$) such that
$\sigma\models\phi$.
We say that two formulas $\phi, \psi \in \LTLP$ are \emph{equivalent on
infinite} (resp., \emph{finite}) \emph{traces}, written $\phi \equiv_I
\psi$ (resp., $\phi \equiv_F \psi$), when, for all $\sigma \in
(2^{\Sigma})^{\omega}$ (resp., $\sigma \in (2^{\Sigma})^{+}$), it holds that
$\sigma$ is a model of $\phi$ if and only if $\sigma$ is a model of $\psi$. We simply
write $\equiv$ when it is clear from the context which one between $\equiv_F$
and $\equiv_I$ has to be used.

If $\phi$ belongs to \LTLFP (\ie pure past fragment of \LTLP), then we interpret $\phi$
only on \emph{finite} state sequences and we say that $\sigma \in
(2^\Sigma)^+$ is a model of $\phi$ if and only if $\sigma,|\sigma|-1 \models \phi$,
that is, each $\phi$ in \LTLFP is interpreted at the \emph{last} state of
a finite state sequence.

\paragraph{Safety and Cosafety Fragments of \LTL.}
We recall the definition of safety and cosafety $\omega$-regular languages.
Let $A$ be a finite alphabet. For any $\sigma \in A^+ \cup A^\omega$ and any $i<|\sigma|$, we denote by $\sigma_{[0,i]}$ the prefix of
$\sigma$ from $0$ to $i$.

\begin{definition}[Co-safety language~\cite{kupferman2001model,thomas1988safety}]
  \label{def:cosafelang}
  Let $\lang \subseteq A^\omega$ (resp., $\lang \subseteq A^+$). We say that
  $\lang$ is a \emph{co-safety language of infinite} (resp., \emph{finite})
  \emph{words} if and only if for all 
  $\sigma \in A^\omega$
  (resp.  $\sigma \in A^+$), it holds that if $\sigma \in \lang$, then
  there exists $i\in\N$ (resp.,  $i<|\sigma|$) such that $\sigma_{[0,i]}\cdot\sigma' \in \lang$, for all
  $\sigma'\in A^\omega$ (resp., $\sigma'\in A^+$). 
\end{definition}


\begin{definition}[Safety language]
  \label{def:safelang}
  A language $\lang$ is a safety language iff its complement
  $\overline{\lang}$ is a cosafety language.
\end{definition}

Let $\mathbb{L}\subseteq\LTLP$. We say that $\mathbb{L}$ is a safety (resp.,
cosafety) fragment of \LTLP iff $\lang(\phi)$ is a safety (resp.,
co-safety) language, for any
$\phi\in\mathbb{L}$.
The following result establishes a connection between the semantic and the
syntactic (co)safety fragment of \LTLP.

\begin{proposition}[\citet{ChangMP92,thomas1988safety,DBLP:conf/fossacs/CimattiGGMT22}]
  \label{prop:galphafalpha}
  Let $\phi$ be a formula of \LTL and let $\lang(\phi)$ be the language of
  $\phi$ over infinite or over finite traces. The following sentences are
  equivalent:
  \begin{itemize}
    \item $\lang(\phi)$ is a safety (resp., co-safety) language;
    \item there exists a formula $\phi'$ in $\Galpha$ (resp.,
      $\Falpha$) such that $\phi'\equiv\phi$;
    \item there exists a formula $\phi''$ in \safetyltl (resp.,
      \cosafetyltl) such that $\phi''\equiv\phi$.
  \end{itemize}
\end{proposition}

\paragraph{Satisfiability and Validity.}
We say that an \LTLP formula $\phi$ is \emph{satisfiable on infinite}
(resp., \emph{finite}) \emph{traces} if there exists a trace $\sigma \in
(2^{\Sigma})^{\omega}$ (resp., $\sigma \in (2^{\Sigma})^{+})$ such that
$\sigma$ is a model of $\phi$.
We say that $\phi$ is \emph{valid on infinite} (resp., \emph{finite})
\emph{traces} if, for every trace $\sigma \in (2^{\Sigma})^{\omega}$
(resp., $\sigma \in (2^{\Sigma})^{+})$, we have that $\sigma$ is a model of
$\phi$.

Given a set of formulas $\mathbb{L}$, the \emph{satisfiability} (resp., \emph{validity}) \emph{problem for $\mathbb{L}$} on finite or
infinite traces, respectively, is the problem of establishing, given a formula $\phi \in
\mathbb{L}$,  whether $\phi$ is satisfiable (resp.,
valid) on infinite or finite traces, respectively.
We recall some results from the literature
on the complexity of the satisfiability and validity problems of
(fragments of) \LTL on infinite and finite traces.

\begin{proposition}[\citet{sistla1985complexity,DeGiacomoV13}]
\label{prop:sat:sistlaclarke}
  The satisfiability problems  for \LTL and
  \LTLP (resp., \LTLxf) on infinite and on finite traces is \PSPACE-complete (resp., \NP-complete).
\end{proposition}

\begin{proposition}[Cf.~\eg~\citet{gabbayetal2003manydimensional}, Section 1.6]
\label{prop:duality}
  Let $\mathbb{L}$ and $\mathbb{L}'$ be two sets of formulas such that $\phi
  \in \mathbb{L}$ iff the transformation into NNF of $\lnot\phi \in
  \mathbb{L}'$, and let $\mathsf{C}$ be a complexity class. It holds that the
  satisfiability problem for $\mathbb{L}$ is $\mathsf{C}$-complete iff the
  validity problem for $\mathbb{L}'$ is $\mathsf{coC}$-complete.
\end{proposition}

From \cref{prop:sat:sistlaclarke,prop:duality}, one can prove the following
result on the complexity of the validity problem for \LTL, \LTLP and
$\LTLwxg$.

\begin{proposition}
\label{prop:val:sistlaclarke}
  The validity problem  for \LTL and \LTLP
  (resp., \LTLwxg) on infinite and on finite traces  is \PSPACE-complete (resp., \coNP-complete).
\end{proposition}

\paragraph{Realizability.}
We define the realizability problem 
for temporal logic formulas
as a two-player game between Controller, whose aim is
to satisfy the formula, and Environment, who tries to violate it. In
this setting, the notion of strategy plays a crucial role.

\begin{definition}[Strategy]
\label{def:strategy}
  Let $\Sigma=\Cset\cup\Uset$ be a set of variables partitioned into
  \emph{controllable} $\Cset$ and \emph{uncontrollable} $\Uset$  ones.
  A \emph{strategy} for \emph{Controller} is a function $s:(2^\Uset)^+ \to
  2^\Cset$ that, for any finite sequence
  $\Uncontr=\seq{\Uncontr_0,\ldots,\Uncontr_n}$ of choices by
  \emph{Environment}, determines the choice $\Contr_n=s(\Uncontr)$ of
  \emph{Controller}.
\end{definition}

Let $s:(2^\Uset)^+ \to 2^\Cset$ be a strategy and let $\Uncontr=\seq{\Uncontr_0,\Uncontr_1,\ldots}$ $\in(2^\Uset)^\omega$ be an infinite sequence of
choices by Environment. We denote
by $\res(s,\Uncontr)= \seq{\Uncontr_0\cup s(\seq{\Uncontr_0}), \Uncontr_1
\cup s(\seq{\Uncontr_0, \Uncontr_1}), \ldots}$ the state sequence resulting
from reacting to $\Uncontr$ according to $s$.  The realizability problem
can be defined as follows.

\begin{definition}[Realizability]
\label{def:realizability}
  Let $\phi$ be
  an $\LTLP$
  formula over the alphabet $\Sigma = \Cset \cup
  \Uset$, with $\Cset\cap\Uset=\emptyset$. We say that $\phi$ is
  \emph{realizable over infinite} (resp., \emph{finite}) \emph{traces} if
  and only if there exists a strategy $s : (2^\Uset)^{+} \to 2^\Cset$ such
  that, for any infinite sequence $\Uncontr=\seq{\Uncontr_0, \Uncontr_1,
  \dots}$ in $(2^\Uset)^\omega$, it holds that $\res(s,\Uncontr) \models
  \phi$ (resp., there exists $k\in\N$ such that the prefix of
  $\res(s,\Uncontr)$ from $0$ to $k$ is a model of $\phi$).
\end{definition}

Given a set of formulas $\mathbb{L}$,
the realizability problem of $\mathbb{L}$ is the problem of establishing, given a formula $\phi\in\mathbb{L}$, whether
$\phi$ is realizable.
We recall some results in the literature on the complexity
of the realizability problem of (fragments of) \LTL and \LTLP over infinite and
finite traces.

\begin{proposition}[\cite{pnueli1989synthesis,rosner1992modular,DeGiacomoV15}]
  Realizability  for \LTL and
  \LTLP over infinite and over finite traces is \EXPTIME[2]-complete.
\end{proposition}


\section{General Results on Finite Traces}
\label{sec:general}

In this section, we provide some theorems that we will use in the following to
determine the complexities of (co)safety fragments interpreted over
finite traces.

We begin with the definition of \emph{suffix independence} for a logic,
which requires infinite models of its formulas to coincide with the
concatenation of finite models with arbitrary infinite traces.
\begin{definition}
\label{def:suffixindep}
  Let $\mathbb{L}$ be a fragment of \LTLP. We say that $\mathbb{L}$ is
  \emph{suffix independent} iff, for any $\phi\in\mathbb{L}$ over the
  alphabet $\Sigma$, $\lang(\phi)=\lang(\phi)_F\cdot(2^\Sigma)^\omega$,
  where $\lang(\phi)$ (resp. $\lang(\phi)_F$) is the language of $\phi$
  over infinite (resp. finite) traces.
\end{definition}

For suffix independent logics, we prove the following equi\-satisfiability
result: if a formula is satisfiable on infinite traces, it is satisfiable also
over finite traces,
and \viceversa.
\nb{A: This is not a Th. just an immediate observation. Remove it}

\begin{restatable}{theorem}{thfininf}
\label{th:fininf}
  Let $\mathbb{L}$ be a fragment of \LTLP that is suffix independent. For
  any $\phi\in\mathbb{L}$, it holds that:
  \begin{align*}
    \lang(\phi) \neq \emptyset \ 
    \Iff \ 
    \lang(\phi)_F \neq \emptyset
  \end{align*}
  where $\lang(\phi)$ (resp. $\lang(\phi)_F$) is the language of $\phi$
  over infinite (resp. finite) traces.
\end{restatable}

The second theorem of this section is a small (bounded) model property for
all safety languages of finite words, which proves that if any of these
languages is not empty, then there is at least a word of length $1$ in the
language.

\begin{restatable}{theorem}{thsmpsafetyfin}
\label{th:smp:safety:fin}
  Let $\lang\subseteq (2^\Sigma)^+$ be a safety language of finite traces. If
  $\lang \neq \emptyset$, then there exists a word
  $\seq{\sigma_0}$ of length $1$ such that $\seq{\sigma_0}\in\lang$.
\end{restatable}

\Cref{th:smp:safety:fin} will let us prove that
the complexity of the satisfiability and realizability problems of safety fragments significantly decreases when passing from infinite to finite words. The next result proves a stronger property of the safety fragments of \LTL. 

\begin{restatable}{theorem}{thsmpsafetyfinlogic}
  \label{thm:smp:safety:fin:logic}
    Let $\mathbb{L} \in \{ \LTLwxg, \safetyltl, \Galpha \}$.
    For any
    $\phi\in\mathbb{L}$
  and any
  $\sigma=\seq{\sigma_0,\dots,\sigma_n}\in(2^\Sigma)^+$ (for some
  $n\in\N$), if $\sigma\models\phi$ then $\seq{\sigma_0}\models\phi$.
\end{restatable}


\section{Complexity of Satisfiability and Validity}
\label{sec:satval}



\begin{table*}[htp]
\centering
  \renewcommand\arraystretch{1.3}%
  \begin{tabular}{ ccccccc }
    \multicolumn{1}{c}{Logics} & \multicolumn{6}{c}{Problems} \\
    \cmidrule(rl){1-1}\cmidrule(rl){2-7}
    & \multicolumn{2}{c}{satisfiability}& \multicolumn{2}{c}{validity}&
      \multicolumn{2}{c}{realizability}\\
    \cmidrule(rl){2-3}\cmidrule(rl){4-5}\cmidrule(rl){6-7}
    & infinite & finite & infinite & finite & infinite & finite \\
   \hline
      \cosafetyltl &\cellcolor{lightgray}\PSPACE-c
                   &\cellcolor{lightgray}\PSPACE-c
                   &\PSPACE-c
                   &\cellcolor{lightgray}\coNP-c
                   &\cellcolor{lightgray}{\EXPTIME[2]-c}
                   &\cellcolor{lightgray}{\EXPTIME[2]-c} \\
      \Falpha &\cellcolor{lightgray}\PSPACE-c
              &\cellcolor{lightgray}\PSPACE-c
              &\cellcolor{lightgray}\PSPACE-c
              &\cellcolor{lightgray}\coNP-c 
              &\cellcolor{lightgray}{\EXPTIME-c} 
              &\cellcolor{lightgray}{\EXPTIME-c} \\
      \LTLxf & \NP-c
             & \NP-c
             &\PSPACE-c 
             &\cellcolor{lightgray}\coNP-c
             &\cellcolor{lightgray}{\EXPTIME-c}
             &\cellcolor{lightgray}{\EXPTIME-c} \\
  \hline
  \end{tabular}
  \caption{Table with complexity results on the cosafety fragments of \LTL (the results in grey are proved in this paper).
  }%
\label{fig:cosafety}
\bigskip
  \begin{tabular}{ ccccccc }
    \multicolumn{1}{c}{Logics} & \multicolumn{6}{c}{Problems} \\
    \cmidrule(rl){1-1}\cmidrule(rl){2-7}
    & \multicolumn{2}{c}{satisfiability}& \multicolumn{2}{c}{validity}&
      \multicolumn{2}{c}{realizability}\\
    \cmidrule(rl){2-3}\cmidrule(rl){4-5}\cmidrule(rl){6-7}
    & infinite & finite & infinite & finite & infinite & finite \\
   \hline
      \safetyltl &\PSPACE-c 
                 &\cellcolor{lightgray}\NP-c 
                 &\cellcolor{lightgray}\PSPACE-c 
                 &\cellcolor{lightgray}\PSPACE-c
                 &\cellcolor{lightgray}{\EXPTIME[2]-c} 
                 &\cellcolor{lightgray}{$\Pi^P_2$-c} \\
      \Galpha &\cellcolor{lightgray}\PSPACE-c 
              &\cellcolor{lightgray}\NP-c
              &\cellcolor{lightgray}\PSPACE-c 
              &\cellcolor{lightgray}\PSPACE-c
              &\cellcolor{lightgray}{\EXPTIME-c} 
              &\cellcolor{lightgray}{$\Pi^P_2$-c} \\
      \LTLwxg &\PSPACE-c
              &\cellcolor{lightgray}\NP-c
              &\coNP-c
              &\coNP-c
              &\cellcolor{lightgray}{\EXPTIME-c} 
              &\cellcolor{lightgray}{$\Pi^P_2$-c} \\
   \hline
  \end{tabular}
  \caption{
Table with complexity results on the safety fragments of \LTL (the results in grey are proved in this paper).
  }%
\label{fig:safety}
\end{table*}

In this section, we study the complexity of the satisfiability and validity
problems for safety and cosafety fragments of \LTLP on both infinite and finite
traces. In particular, here and in the rest of the paper, we will focus on the
cosafety fragments \cosafetyltl, \Falpha, and \LTLxf, and the dual safety
fragments \safetyltl, \Galpha, and \LTLwxg. On infinite traces, we show the
following results.

\begin{theorem}
\label{thm:inf:satval}
The satisfiability and validity problems on infinite traces are
\PSPACE-complete for:
\begin{enumerate}
	\item \cosafetyltl, \Falpha;
	\item \LTLwxg, \safetyltl, \Galpha.
\end{enumerate}
\end{theorem}
Moreover, we prove the following results on finite traces.
\begin{theorem}
\label{thm:fin:satval}
  The satisfiability problem on finite traces is:
  \begin{enumerate}
  	\item \PSPACE-complete for \cosafetyltl, \Falpha;
    \item \NP-complete for \LTLwxg, \safetyltl, \Galpha.
  \end{enumerate}
  The validity problem on finite traces is \coNP-complete for \LTLxf,
  \cosafetyltl, \Falpha.
\end{theorem}

The results stated by \cref{thm:inf:satval,thm:fin:satval}, which are summarised
in~\cref{fig:cosafety,fig:safety}, show a surprising 
(a)symmetry in the complexity of the satisfiability problem along two different dimensions.

Moving \emph{from \cosafetyltl to \LTLxf}, either on infinite or on finite
traces, the complexity of satisfiability changes from \PSPACE-complete to
\NP-complete.  This comes from a linear-size model property known for \LTLxf on infinite traces~\cite[Lemma 3.6]{sistla1985complexity}, which allows us to
guess (nondeterministically) a candidate model and then check it in polynomial
time. Instead, thanks to the \emph{until} ($\ltl{U}$) operator in \cosafetyltl,
which combines an existential quantification over time points with a bounded
universal one, we are able to encode \LTL formulas interpreted over finite
traces with \cosafetyltl formulas that, by means of the \emph{until} modality,
can hook the final state of a finite trace and simulate the universal temporal
modalities of \LTL (like the \emph{globally}) by means of the universal part of
the \emph{until}.

It is worth noticing that, being without universal temporal operators (that is, $\ltl{wX}$, $\ltl{G}$, and $\ltl{R}$), \cosafetyltl, \Falpha, and \LTLxf
formulas cannot detect 
any difference between satisfiability on finite and on infinite traces, since
any satisfying finite trace can be arbitrarily extended to an infinite model,
and any satisfying infinite trace can be suitably contracted to a satisfying
finite prefix. In fact, we will prove that \cosafetyltl, \LTLxf, and \Falpha are
suffix independent logics (\cref{def:suffixindep}), and we will use
\cref{th:fininf} to prove that their complexities do not change when considering
finite or infinite traces.

In contrast to cosafety fragments, the complexity of safety fragments
significantly changes going from infinite to finite traces: while satisfiability
is \PSPACE-complete on infinite traces, it is \NP-complete on finite traces.
This is because the \emph{weak next} ($\ltl{wX}$) operator, available in
\safetyltl and \LTLwxg, behaves on infinite traces exactly as a \emph{strong
next} ($\ltl{X}$), which, together with the \emph{globally} ($\ltl{G}$) or the
\emph{release} ($\ltl{R}$) operators, can encode computations of Turing machines
with a polynomial tape \cite[cf.][Theorem~15.8.1]{gabbay1994temporal}. Instead,
on finite traces, the combination of $\ltl{wX}$ and $\ltl{G}$ cannot force a
trace to have more then one state. In fact, for any safety fragment interpreted
over finite traces, by \cref{th:smp:safety:fin}, we have that any formula of
these logics is satisfiable if and only if it has a model of length~$1$, which leads to the \NP complexity.

\cref{thm:inf:satval,thm:fin:satval} are proved in
the rest of this section.

\subsection{Complexity on Infinite Traces}

We begin with the proof of
\cref{thm:inf:satval},
proving first
the \PSPACE-completeness of satisfiability on infinite traces for
\cosafetyltl and \Falpha.

We start from \cosafetyltl.
To prove \PSPACE-hardness, we reduce
the satisfiability problem of \LTL over finite traces, which is
\PSPACE-complete~\cite{DeGiacomoV13}, to the satisfiability of
\cosafetyltl over infinite traces.
For any formula $\phi \in \LTL$, we will define a formula $g(\phi) \in
\cosafetyltl$ such that:
\begin{enumerate*}[label=(\roman*)]
  \item the size of $g(\phi)$ is polynomial in the size of $\phi$;
  \item $\phi$ is satisfiable over finite traces if and only if $g(\phi)$ is
    satisfiable over infinite traces.
\end{enumerate*}
The rationale is to introduce a fresh proposition letter $\endt$ that is
supposed to represent, in an infinite state sequence, the end of a finite
trace.
We first define a transformation $f(\cdot)$ from \LTL to \cosafetyltl
formulas inductively as follows:
\nb{M: changed, to check}
\begin{align*}
  f(p) \coloneqq {} & p,\quad \text{for any $p\in\Sigma$},\\
  f(\lnot p) \coloneqq {} & \lnot p,\quad \text{for any $p\in\Sigma$},\\
  f(\phi_1 \land \phi_2) \coloneqq {} & f(\phi_1) \land f(\phi_2),\\
  f(\phi_1 \lor \phi_2) \coloneqq {} & f(\phi_1) \lor f(\phi_2),\\
  f(\ltl{X\phi_1}) \coloneqq {} & \ltl{X(! \endt && f(\phi_1))},\\
  f(\ltl{wX\phi_1}) \coloneqq {} & \ltl{X(! \endt -> f(\phi_1))},\\
  f(\ltl{\phi_1 U \phi_2}) \coloneqq {} & \ltl{(! \endt \land f(\phi_1) ) U (! \endt
    && f(\phi_2))},\\
  f(\ltl{\phi_1 R \phi_2}) \coloneqq {} & \ltl{( (! \endt \land f(\phi_2) ) U (! \endt && f(\phi_2) && X(\endt)))} \ \lor \\
        & \ltl{( (! \endt \land f(\phi_2) ) U (! \endt && f(\phi_1) && f(\phi_2))}).
\end{align*}
On finite and infinite traces,
$f(\ltl{F\phi_1})$ and $f(\ltl{G\phi_1})$ can be equivalently rewritten as
$\ltl{( ! \endt ) U ( ! \endt && f(\phi_1) )}$,
and
$\ltl{( ! \endt && f(\phi_1) ) U( ! \endt && f(\phi_1) && X(\endt))}$,
respectively.
Starting from $f(\cdot)$, we define the transformation $g(\cdot) : \LTL \to
\cosafetyltl$ as follows: for any $\phi\in\LTL$, we define $g(\phi)
\coloneqq \ltl{! \endt && f(\phi) && (! \endt) U (\endt)}$.  For any
$\phi\in\LTL$, $g(\phi)$ is a \cosafetyltl formula and the size of
$g(\phi)$ is polynomial (more precisely, linear) in the size of $\phi$.
The following lemma establishes the main property for using $g(\cdot)$ as
an hardness reduction.

\begin{restatable}{lemma}{lemmasatendt}
\label{lemma:sat:endt}
  For any $\phi\in\LTL$, it holds that $\phi$ is satisfiable over finite
  traces iff $g(\phi)$ is satisfiable over infinite traces.
\end{restatable}

Using \cref{lemma:sat:endt}, we can easily prove the following result.

\begin{lemma}
\label{sat:cosafetyltl}
  The satisfiability problem for \cosafetyltl on infinite traces is
  \PSPACE-complete.
\end{lemma}
\begin{proof}
  \emph{(Membership)} Immediate from the fact that $\cosafetyltl \subseteq
  \LTL$ and that \LTL satisfiability on infinite traces is
  \PSPACE-complete~\cite{sistla1985complexity}.

  \emph{(Hardness)} Immediate from \cref{lemma:sat:endt} and the fact that
  the size of $g(\phi)$ is polynomial in the size of $\phi$.
\end{proof}

We now prove the \PSPACE-completeness for the satisfiability problem over
infinite traces of the \Falpha fragment. The hardness proof is based on the
simple consideration that any formula $\phi$ of \LTLFP is satisfiable (over
finite traces) if and only if the formula $\ltl{F}(\phi)$ is satisfiable over infinite
(or finite) traces. The \PSPACE-hardness follows from the fact that
satisfiability of \LTLFP is \PSPACE-complete.\footnotemark

\footnotetext{
To see this, observe
  that any formula $\phi$ of
  \LTL is satisfiable over finite traces iff $\phi'$ is satisfiable, where
  $\phi'$ is obtained from $\phi$ by replacing each $\ltl{X}$ (resp.
  $\ltl{U}$) operator with $\ltl{Y}$
  (resp. $\ltl{S}$).
  The
  \PSPACE-completeness follows from the fact that satisfiability over
  finite traces is \PSPACE-complete~\cite{DeGiacomoV13}.
}

\begin{restatable}{lemma}{lemmasatfalpha}
\label{sat:falpha}
  The satisfiability problem for \Falpha on infinite traces is
  \PSPACE-complete.
\end{restatable}

From \cref{sat:cosafetyltl,sat:falpha,prop:duality}, it follows that the
validity problem of \safetyltl and \Galpha over infinite traces is
\PSPACE-complete.


We now focus on
complexity of satisfiability for \safetyltl,
\Galpha, and \LTLwxg on infinite traces,
showing that
all
these
problem are also \PSPACE-complete.

\PSPACE-completeness of \LTLwxg follows from the same proof as
\cite[Th. 5.2 Cor. 5.1]{cimatti2021extended} or, alternatively, by adapting
the proof by ~\cite[Thm.~15.8.1]{gabbay1994temporal} or the proof by
\cite[Thm.~4.4]{artaleetal2014cookbook}.
\PSPACE-completeness of the validity
problem for \Falpha over infinite traces follows from \cref{prop:duality}.  Moreover, since
\LTLwxg is a syntactic fragment of \safetyltl, it immediately follows
that satisfiability (resp., validity) of \safetyltl (resp., \cosafetyltl)
over infinite traces is \PSPACE-complete.

Finally, we have to prove that satisfiability of \Galpha is \PSPACE-complete on infinite traces.  To prove it, we show that
the validity problem for \Falpha is \PSPACE-complete: \PSPACE-completeness of satisfiability of \Galpha
then follows from
\cref{prop:duality}. As in \cref{sat:falpha}, the validity of \Falpha can be reduced
to the validity of \LTLFP.

\begin{restatable}{lemma}{valinffalpha}
\label{val:inf:falpha}
  The validity problem for \Falpha on infinite traces  is \PSPACE-complete.
\end{restatable}

From \cref{val:inf:falpha,prop:duality},
it
follows that
\Galpha
satisfiability
on
infinite traces  is \PSPACE-complete.

\subsection{Complexity on Finite Traces}

We now move to the proof of \cref{thm:fin:satval}.  We first show the
\PSPACE-completeness of the satisfiability (resp.  validity) problem of
\cosafetyltl and \Falpha (resp. \safetyltl and \Galpha) over finite traces. To
this goal, we first prove that \cosafetyltl and \Falpha are suffix
independent. We will use this result, along with \cref{th:fininf}, to transfer
the complexity of satisfiability from infinite to finite traces (cf.
also~\citet{DBLP:conf/fossacs/CimattiGGMT22}, Lemma 1,
and~\citet{artale22TOCLarxivsubmission}, Lemma 4.11).

\begin{restatable}{lemma}{lemfininflogics}
\label{lemma:fininf:logics}
  \cosafetyltl and \Falpha are suffix independent.
\end{restatable}

From \cref{th:fininf}, we obtain the following corollary.
\begin{corollary}
\label{cor:sat:cosafety:fin}
  The satisfiability problem over finite traces of \cosafetyltl and \Falpha
  is \PSPACE-complete.
\end{corollary}

By \cref{prop:duality}, it follows that the validity problems of
\safetyltl and \Galpha are \PSPACE-complete.

It is worth noticing that \cref{th:fininf} does not work for safety
fragments of \LTL: for example, the formula $\ltl{G(wX \bot)}$ is
satisfiable over finite traces but unsatisfiable over infinite traces. As
a matter of fact, below we show that the complexity of \safetyltl,
\Galpha and \LTLwxg satisfiability lowers down to \NP-complete under
finite trace semantics. Indeed, since \safetyltl, \Galpha and \LTLwxg are
safety fragments of \LTL, from \cref{th:smp:safety:fin}, it follows that
any satisfiable formula of these fragments has a model of length~1.


Consequently, we can give a nondeterministic algorithm that,
in polynomial time, solves the satisfiability of a formula
$\phi\in\mathbb{L}$,
with
$\mathbb{L} \in \{ \LTLwxg, \safetyltl, \Galpha \}$.
It simply suffices to guess an
assignment for the initial state of a candidate trace and check if it
satisfies $\phi$. If such an assignment is found, then it means that $\phi$
is satisfiable, otherwise, by \cref{th:smp:safety:fin}, $\phi$ is
unsatisfiable. This proves the membership of \safetyltl, \LTLwxg, and
\Galpha to \NP. The hardness simply follows from a reduction of the SAT
problem.

\begin{restatable}{lemma}{corsmpsafety}
\label{sat:fin:safetyltl:ltlwxg:galpha}
  The satisfiability problem on finite traces for \safetyltl, \LTLwxg,
  and \Galpha is \NP-complete.
\end{restatable}

By \cref{prop:duality},
the validity problem
on
finite
traces for \cosafetyltl, \LTLxf, and \Falpha is \coNP-complete.


\section{Complexity of Realizability}
\label{sec:real}

In this section, we study the complexity of the realizability problem for
the (co)safety fragments of \LTL that we considered in the previous
section. The following theorems sum up our results on realizability over
infinite and finite traces.

\begin{theorem}
\label{thm:inf:real}
  The realizability problem over infinite traces is
  \begin{itemize}
    \item \EXPTIME[2]-complete for \safetyltl, \cosafetyltl;
    \item \EXPTIME-complete for \Galpha, \Falpha, \LTLwxg, and \LTLxf.
  \end{itemize}
\end{theorem}

\begin{theorem}
\label{thm:fin:real}
  The realizability problem over finite traces is
  \begin{itemize}
    \item \EXPTIME[2]-complete for \cosafetyltl;
    \item \EXPTIME-complete for \Falpha, \LTLxf;
    \item $\Pi^P_2$-complete for \safetyltl, \Galpha, and \LTLwxg.
  \end{itemize}
\end{theorem}

\subsection{Complexity on Infinite Traces}

We first prove the \EXPTIME[2]-completeness of \cosafetyltl realizability
on infinite (and finite) traces. To show hardness, we consider
the realizability problem of \LTL over finite traces, which is
\EXPTIME[2]-complete~\cite{DeGiacomoV15}. For any formula $\phi$ of \LTL,
we consider the formula $g(\phi)$ as defined in the previous section, and
we define the uncontrollable variable $\Uset'$ of $g(\phi)$ as the
uncontrollable variables of $\phi$, and the controllable variables $\Cset'$
of $g(\phi)$ as the set of controllable variables of $\phi$ and $\endt$.
The following lemma establishes the equirealizability between $\phi$ (over
finite traces) and $g(\phi)$ over infinite traces.

\begin{restatable}{lemma}{lemmarealendt}
\label{lemma:real:endt}
  For any $\phi\in\LTL$, it holds that $\phi$ is realizable over finite
  traces iff $g(\phi)$ is realizable over infinite traces.
\end{restatable}

We use \cref{lemma:real:endt} as the core of a reduction from
realizability of \LTL over infinite traces to realizability of \cosafetyltl
over finite traces,
thus proving the following.

\begin{restatable}{lemma}{realcosafetyltl}
\label{real:cosafetyltl}
  The realizability problem over infinite traces for \cosafetyltl is
  \EXPTIME[2]-complete.
\end{restatable}

We now study the complexity of \LTLwxg and \Galpha. Interestingly, for
these two fragments the realizability problem over infinite traces is
\EXPTIME-complete. In fact, as described
in~\cite{de2021pure,cimatti2021extended}, for any formula $\phi$ in \LTLwxg
or in \Galpha, there exists (and can be actually built effectively)
a deterministic finite automaton (\DFA) $\autom(\lnot\phi)$ such that:
\begin{enumerate*}[label=(\roman*)]
  \item its language is exactly the set of bad prefixes of $\phi$; and
  \item its size is singly exponential in the size of $\phi$.\footnotemark
\end{enumerate*}
\footnotetext{
  We recall that, in the general case, the construction of a \DFA starting
  from an \LTL formula interpreted over finite traces requires two steps,
  each introducing an exponential blowup in the worst case:
  \begin{enumerate*}[label=(\roman*)]
    \item the transformation of the \LTL formula into a
    non-deterministic finite automaton
    (\NFA);
    \item the determinization of the \NFA through the classic subset
      construction.
  \end{enumerate*}
}
Then, realizability can be solved on top
of $\autom(\lnot\phi)$ by checking whether Controller can force the game to
never visit a final state of the automaton. This kind of games, called
\emph{safety games}, can be solved in linear time. It follows that \Galpha
and \LTLwxg realizability (over infinite traces) belongs to \EXPTIME.


The \EXPTIME-hardness of \LTLwxg follows from~\cite[Th. 5.2,
Cor.5.1]{cimatti2021extended}.
The \EXPTIME-hardness of \Galpha realizability over infinite words can be
proved in a similar way as for the \LTLwxg case: for any infinite corridor
tiling game $\tiling$~\cite{chlebus1986domino}, we build a corresponding
\LTLwxg formula $\phi$ such that $\tiling$ admits a strategy iff $\phi$ is
realizable. It is worth noticing that this encoding can be derived from the
one of \LTLEBRP with no bounded operators~\cite{cimatti2021extended} by
using the $\ltl{Y}$ operators instead of $\ltl{X}$.

\begin{restatable}{lemma}{realgalpha}
\label{real:galpha}
  The realizability problem over infinite traces of \LTLwxg and \Galpha is
  \EXPTIME-complete.
\end{restatable}

We now prove a lemma that allows us to dualize the complexities for
realizability (over infinite traces) we have found so far for \cosafetyltl,
\LTLwxg, and \Galpha to \safetyltl, \LTLxf, and \Falpha, respectively.  The
following lemma can be considered as the version of \cref{prop:duality} for
realizability.

\begin{restatable}{lemma}{realduality}
\label{real:duality}
  Let $\mathbb{L}$ be \cosafetyltl (resp. \LTLxf, resp. \Falpha) and let
  $\mathbb{L}'$ be \safetyltl (resp. \LTLwxg, resp. \Galpha).
For a complexity class  $\mathsf{C}$, 
  the realizability problem
  over
  infinite traces for
  $\mathbb{L}$ is $\mathsf{C}$-complete iff the realizability problem over
  infinite traces for $\mathbb{L}'$ is $\mathsf{coC}$-complete.
\end{restatable}

The rationale behind \cref{real:duality} is that realizability games are
zero-sum games~\cite{jacobs2017first}%
: Controller has a winning strategy for $\phi$ iff
Environment has not a winning strategy for $\lnot\phi$. Crucially, the
existence of a winning strategy of Environment for $\lnot\phi$ can be
checked with classical realizability: it suffices to swap the controllable
variables of $\lnot\phi$ with the uncontrollable ones, and \viceversa, and
to codify in the formula the fact that Environment player has to play as
the second player. \Cref{real:duality}, together with
\cref{real:cosafetyltl,real:galpha}, implies the following complexity
results\footnotemark.

\begin{lemma}
\label{real:safetyltl}
  The realizability problem over infinite traces for $\safetyltl$ (resp.
  $\Falpha$ and $\LTLxf$) is \EXPTIME[2]-complete (resp.
  \EXPTIME-complete).
\end{lemma}

\footnotetext{
  Note that this contradicts~\citet{arteche2021complexity}, who
  acknowledged a flaw in their article.
}

\subsection{Complexity on Finite Traces}

It is simple to see that \cref{th:fininf} implies that, for any formula
$\phi$ of \cosafetyltl, \Falpha or \LTLxf, $\phi$ is realizable over
infinite traces iff $\phi$ is realizable over finite traces. Therefore, we
have that the realizability problem over finite traces of \cosafetyltl,
\Falpha and \LTLxf is \EXPTIME[2]-complete, \EXPTIME-complete, and
\EXPTIME-complete, respectively.

We prove that, similarly for the case of satisfiability, the complexity of
safety fragments for the realizability problem significantly decreases when
passing from infinite to finite traces. In particular, we prove that
realizability over finite traces of \LTLwxg, \safetyltl and \Galpha is
$\Pi^P_2$-complete.  We first prove the following small model property
(analogous to \cref{th:smp:safety:fin} for satisfiability), which follows from
\cref{thm:smp:safety:fin:logic}.

\begin{restatable}{lemma}{thmrealcontraction}
\label{real:contraction}
  Let
  $\mathbb{L} \in \{ \LTLwxg, \safetyltl, \Galpha \}$. Any
    $\phi\in\mathbb{L}$ is realizable
  on finite traces iff there exists a strategy $s : (2^\Uset)^+ \to
  (2^\Cset)$ such that $\res(s,\Uncontr)_0 \models \phi$, for any
  $\Uncontr\in(2^\Uset)^\omega$.
\end{restatable}

\Cref{real:contraction} allows for the following algorithm deciding the
realizability over finite traces of \LTLwxg, \safetyltl and \Galpha: for
any $\phi$ in these fragments, check the existence of a strategy that
satisfies $\phi$ in one step; if it exists, $\phi$ is realizable;
otherwise, by \cref{real:contraction}, it is unrealizable.

The existence of a strategy implementing $\phi$ in one step amounts to the check
of satisfiability of a \emph{Quantified Boolean Formula} with one quantifier
alternation (2QBF), which is a $\Pi^P_2$-complete problem~\cite{BuningB09}. In
the following we describe the algorithm.

We start with
$\mathbb{L} \in \{ \LTLwxg, \safetyltl \}$.
Let
$\phi \in \mathbb{L}$
and let
$\Uset=\set{u_1,\dots,u_m}$ (resp.  $\Cset=\set{c_1,\dots,c_n}$) be the set of
uncontrollable (resp. controllable) variables of $\phi$.
\begin{enumerate}
  \item expand the temporal operators of $\phi$ in the classical fashion
    ($\ltl{G\phi}$ is expanded in $\ltl{\phi \land wX G \phi}$ and
    $\ltl{\phi_1 R \phi_2}$ is expanded in $\ltl{(\phi_1 \land \phi_2) \lor
    (\phi_2 \land wX(\phi_1 R \phi_2))}$); the formula obtained in this way
    is a Boolean combination of proposition atoms or formulas of type
    $\ltl{wX \phi}$;
  \item replace each formula of type $\ltl{wX}\phi$ with $\top$; the
    resulting formula, that we call $\phi'$, is a Boolean formula;
  \item check the satisfiability of  $\forall u_1 \dots \forall u_m \exists c_1
    \dots \exists c_n \suchdot \phi'$, which is a 2QBF formula.
\end{enumerate}

For $\phi\in\Galpha$, the method is the same: for any $\phi$ of type
$\ltl{G(\alpha)}$, we drop the $\ltl{G}$ operator, we expand the past temporal
operators in $\alpha$ and we replace each subformula of type $\ltl{wY\phi}$
(resp. $\ltl{Y\phi}$) with $\top$ (resp. $\bot$).  This gives us the
$\Pi^p_2$-membership of \LTLwxg, \safetyltl and \Galpha realizability over
finite traces. The $\Pi^P_2$-hardness comes directly from the $\Pi^P_2$-hardness
of 2QBF.

\begin{lemma}
\label{real:fin:safety}
  The realizability problem over finite traces of \LTLwxg, \safetyltl and
  \Galpha is $\Pi^P_2$-complete.
\end{lemma}


\section{Discussion}
\label{sec:discussion}

The
complexity gap of satisfiability
for the safety fragments
when moving from infinite to finite traces
is worth discussing. To some
extent, this shows that reducing the problem to considering prefixes of an
$\omega$-language, in the worst case, does not affect the complexity (in fact,
on infinite traces,
the satisfiability problem for all the fragments, except
\LTLxf, has
the same complexity as for full \LTL). On the contrary,
considering the prefixes of a language of finite words can dramatically
decrease the complexity.\fitpar

In the case of infinite trace semantics, in contrast to what happens for
satisfiability, considering safety properties can decrease the worst-case
complexity of realizability with respect to full \LTL
(\cref{real:safetyltl,real:galpha}).
This is due to the crucial role that
\emph{determinism} has in realizability.
Indeed,
realizability is (almost
always) solved by playing a game over an automaton, also called \emph{arena},
whose solution requires a deterministic representation of the arena.
Therefore,
reducing to reasoning over finite words (the main advantage of
considering (co)safety properties) can be exploited by realizability
algorithms,
e.g.
by building a deterministic automaton for 
a $\LTLFP$
formula with only single exponential blowup.  On the contrary,
satisfiability is not able to exploit determinism to improve
worst-case
complexity,
since it
can be solved simply as the reachability of a final
state in a (possibly nondeterministic) automaton corresponding to the
formula.
In other words, determinization is not necessary for satisfiability, and indeed
\cosafetyltl and \Falpha share the same complexity for satisfiability.

Consider now the difference between the complexity of realizability of
\cosafetyltl and \Falpha (or equivalently of \safetyltl and \Galpha).
Despite having the same expressive power (recall \cref{prop:galphafalpha}),
the complexity is significantly lower if the formula is given in the form
\Falpha. This difference has one of these two consequences:
\begin{itemize}
  \item either \cosafetyltl can be exponentially more succinct than
    \Falpha, \ie there exists a formula $\phi\in\cosafetyltl$ such that,
    for all $\phi'\in\Falpha$, if $\phi\equiv_I\phi'$ then
    $\phi'\in\mathcal{O}(2^{|\phi|})$;
  \item or there exists an algorithm of exponential running time such that,
    given any $\phi\in\cosafetyltl$, outputs an equivalent formula $\phi'\in\Falpha$
    with
    $|\phi'|\in\mathcal{O}(\mathsf{poly}(|\phi|))$.
\end{itemize}
Clearly, exactly one of the two points can be true. We conjecture the first
one to be true, but the question is still open.

As already noted in
\cite{de2021pure},
results on computation tree logic
and alternating-time temporal logic
satisfiability~\cite{kupferman2012once,bozzelli2020alt}
could be adapted to show the \EXPTIME-membership of realizability for \Falpha.
It is unclear, however,
how to
use these results
to address
the \EXPTIME lower bound.
We also remark that our result on the \EXPTIME[2]-completeness of
\safetyltl realizability shows the optimality of the algorithm
in~\cite{ZhuTLPV17}.

\citet{FiondaG18} study the complexity of satisfiability for fragments of \LTL over
finite traces, with $\ltl{X}$, $\ltl{F}$ and $\ltl{G}$ as the only available temporal modalities,
by imposing
several
syntactical restrictions and proving a linear-length model
property for some of such fragments.
Our study considers
(together with $\ltl{U}$ and $\ltl{R}$)
also the role
of the $\ltl{wX}$ operator, which is crucial when negation is applied only
to propositional atoms. In addition, we prove that for \emph{all} safety
languages of finite words, there is a \emph{constant-size model property},
allowing one to consider only the first state of a model.


\section{Conclusions}
\label{sec:conclusion}

In this paper, we studied the complexity of the (co)safety fragment of \LTL
for the problems of satisfiability, validity, and realizability, both over
infinite and finite trace semantics.  In particular, we considered three
cosafety fragments (\cosafetyltl, \Falpha, and \LTLxf) and their dual
safety fragments (\safetyltl, \Galpha, and \LTLwxg).

Our results show that:
\begin{enumerate*}[label=(\roman*)]
  \item for the cosafety fragment, the complexities never change when
    passing from infinite to finite trace semantics;
  \item on the contrary, for the safety fragment, considering finite trace
    semantics can significantly decrease the complexity of both
    satisfiability and realizability;
  \item for realizability, past operators play a crucial role; \eg
    by using the \Galpha fragment one can solve realizability in singly
    exponential time while being able to express all safety properties
    definable in \LTL.
\end{enumerate*}

Model-checking is central in the field of temporal logic. A careful
analysis of its complexity for the fragments that we considered in this
paper is 
an interesting
future development.

Finally, our conjecture that \cosafetyltl can be exponentially more succinct than
\Falpha surely deserves an answer. More generally, a careful study of the
succinctness of all fragments (in particular the ones that are expressively
equivalent) seems a promising direction.

\paragraph{Acknowledgments.}
Nicola Gigante acknowledges the support of the PURPLE project, in the
context of the AIPlan4EU project's First Open Call for Innovators.


\bibliography{biblio}

\begin{thebibliography}{38}
\providecommand{\natexlab}[1]{#1}

\bibitem[{Artale et~al.(2014)Artale, Kontchakov, Ryzhikov, and
  Zakharyaschev}]{artaleetal2014cookbook}
Artale, A.; Kontchakov, R.; Ryzhikov, V.; and Zakharyaschev, M. 2014.
\newblock A Cookbook for Temporal Conceptual Data Modelling with Description
  Logics.
\newblock \emph{{ACM} Trans. Comput. Log.}, 15(3): 25:1--25:50.

\bibitem[{Artale, Mazzullo, and Ozaki(2019)}]{AMO:IJCAI19}
Artale, A.; Mazzullo, A.; and Ozaki, A. 2019.
\newblock Do You Need Infinite Time?
\newblock In \emph{Proceedings of the 28th International Joint Conference on
  Artificial Intelligence (IJCAI-19)}. AAAI Press.

\bibitem[{Artale, Mazzullo, and Ozaki(2022)}]{artale22TOCLarxivsubmission}
Artale, A.; Mazzullo, A.; and Ozaki, A. 2022.
\newblock First-order Temporal Logic on Finite Traces: Semantic Properties,
  Decidable Fragments, and Applications.
\newblock \emph{arXiv preprint}, abs/2202.00610.

\bibitem[{Arteche and Hermo(2021)}]{arteche2021complexity}
Arteche, N.; and Hermo, M. 2021.
\newblock On the Complexity of Realizability for Safety LTL and Related
  Subfragments.
\newblock \emph{arXiv preprint}, abs/2112.14102.

\bibitem[{Arteche and Hermo(2024)}]{arteche2024towardsexact}
Arteche, N.; and Hermo, M. 2024.
\newblock Towards the exact complexity of realizability for Safety {LTL}.
\newblock \emph{J. Log. Algebraic Methods Program.}, 141: 101002.

\bibitem[{Bozzelli, Murano, and Sorrentino(2020)}]{bozzelli2020alt}
Bozzelli, L.; Murano, A.; and Sorrentino, L. 2020.
\newblock Alternating-time temporal logics with linear past.
\newblock \emph{Theor. Comput. Sci.}, 813: 199--217.

\bibitem[{Chang, Manna, and Pnueli(1992)}]{ChangMP92}
Chang, E.~Y.; Manna, Z.; and Pnueli, A. 1992.
\newblock Characterization of Temporal Property Classes.
\newblock In Kuich, W., ed., \emph{Proceedings of the 19th International
  Colloquium on Automata, Languages and Programming (ICALP-92)}, 474--486.
  Springer.

\bibitem[{Chlebus(1986)}]{chlebus1986domino}
Chlebus, B.~S. 1986.
\newblock Domino-tiling games.
\newblock \emph{Journal of Computer and System Sciences}, 32(3): 374--392.

\bibitem[{Cimatti et~al.(2021)Cimatti, Geatti, Gigante, Montanari, and
  Tonetta}]{cimatti2021extended}
Cimatti, A.; Geatti, L.; Gigante, N.; Montanari, A.; and Tonetta, S. 2021.
\newblock Extended bounded response LTL: {A} new safety fragment for efficient
  reactive synthesis.
\newblock \emph{Formal Methods in System Design}, 1--49.

\bibitem[{Cimatti et~al.(2022)Cimatti, Geatti, Gigante, Montanari, and
  Tonetta}]{DBLP:conf/fossacs/CimattiGGMT22}
Cimatti, A.; Geatti, L.; Gigante, N.; Montanari, A.; and Tonetta, S. 2022.
\newblock A first-order logic characterisation of safety and co-safety
  languages.
\newblock In \emph{Proceedings of the 25th International Conference on
  Foundations of Software Science and Computation Structures ({FOSSACS}-22)},
  244--263. Springer.

\bibitem[{{De Giacomo}, {De Masellis}, and Montali(2014)}]{DegEtAl}
{De Giacomo}, G.; {De Masellis}, R.; and Montali, M. 2014.
\newblock Reasoning on {LTL} on Finite Traces: Insensitivity to Infiniteness.
\newblock In \emph{Proceedings of the 28th National Conference on Artificial
  Intelligence {(AAAI-14)}}, 1027--1033. {AAAI} Press.

\bibitem[{De~Giacomo et~al.(2021)De~Giacomo, Di~Stasio, Fuggitti, and
  Rubin}]{de2021pure}
De~Giacomo, G.; Di~Stasio, A.; Fuggitti, F.; and Rubin, S. 2021.
\newblock Pure-past linear temporal and dynamic logic on finite traces.
\newblock In \emph{Proceedings of the 29th International Conference on
  International Joint Conferences on Artificial Intelligence (IJCAI-21)},
  4959--4965.

\bibitem[{{De Giacomo} and Vardi(2013)}]{DeGiacomoV13}
{De Giacomo}, G.; and Vardi, M.~Y. 2013.
\newblock Linear Temporal Logic and Linear Dynamic Logic on Finite Traces.
\newblock In Rossi, F., ed., \emph{Proceedings of the 23rd International Joint
  Conference on Artificial Intelligence (IJCAI-13)}, 854--860. {IJCAI/AAAI}.

\bibitem[{{De Giacomo} and Vardi(2015)}]{DeGiacomoV15}
{De Giacomo}, G.; and Vardi, M.~Y. 2015.
\newblock Synthesis for {LTL} and {LDL} on Finite Traces.
\newblock In Yang, Q.; and Wooldridge, M.~J., eds., \emph{Proceedings of the
  24th International Joint Conference on Artificial Intelligence (IJCAI-15)},
  1558--1564. {AAAI} Press.

\bibitem[{Ehlers(2013)}]{ehlers2013symmetric}
Ehlers, R. 2013.
\newblock \emph{Symmetric and efficient synthesis}.
\newblock Ph.D. thesis, Universit{\"{a}}t Saarbr{\"{u}}cken.

\bibitem[{Fionda and Greco(2018)}]{FiondaG18}
Fionda, V.; and Greco, G. 2018.
\newblock {LTL} on Finite and Process Traces: Complexity Results and a
  Practical Reasoner.
\newblock \emph{J. Artif. Intell. Res.}, 63: 557--623.

\bibitem[{Gabbay, Hodkinson, and Reynolds(1994)}]{gabbay1994temporal}
Gabbay, D.~M.; Hodkinson, I.; and Reynolds, M.~A. 1994.
\newblock \emph{Temporal logic: Mathematical Foundations and Computational
  Aspects}, volume~1.
\newblock Clarendon Press.

\bibitem[{Gabbay et~al.(2003)Gabbay, Kurucz, Wolter, and
  Zakharyaschev}]{gabbayetal2003manydimensional}
Gabbay, D.~M.; Kurucz, A.; Wolter, F.; and Zakharyaschev, M. 2003.
\newblock \emph{Many-dimensional Modal Logics: Theory and Applications}, volume
  148 of \emph{Studies in Logic and The Foundations of Mathematics}.
\newblock Elsevier.

\bibitem[{Jacobs et~al.(2017)Jacobs, Bloem, Brenguier, Ehlers, Hell,
  K{\"{o}}nighofer, P{\'{e}}rez, Raskin, Ryzhyk, Sankur, Seidl, Tentrup, and
  Walker}]{jacobs2017first}
Jacobs, S.; Bloem, R.; Brenguier, R.; Ehlers, R.; Hell, T.; K{\"{o}}nighofer,
  R.; P{\'{e}}rez, G.~A.; Raskin, J.; Ryzhyk, L.; Sankur, O.; Seidl, M.;
  Tentrup, L.; and Walker, A. 2017.
\newblock The first reactive synthesis competition {(SYNTCOMP}-14).
\newblock \emph{Int. J. Softw. Tools Technol. Transf.}, 19(3): 367--390.

\bibitem[{Kleine{ }B{\"{u}}ning and Bubeck(2009)}]{BuningB09}
Kleine{ }B{\"{u}}ning, H.; and Bubeck, U. 2009.
\newblock Theory of Quantified Boolean Formulas.
\newblock In \emph{Handbook of Satisfiability}, volume 185 of \emph{Frontiers
  in Artificial Intelligence and Applications}, 735--760. {IOS} Press.

\bibitem[{Kupferman, Pnueli, and Vardi(2012)}]{kupferman2012once}
Kupferman, O.; Pnueli, A.; and Vardi, M.~Y. 2012.
\newblock Once and for all.
\newblock \emph{J. Comput. Syst. Sci.}, 78(3): 981--996.

\bibitem[{Kupferman and Vardi(2001)}]{kupferman2001model}
Kupferman, O.; and Vardi, M.~Y. 2001.
\newblock Model checking of safety properties.
\newblock \emph{Formal Methods in System Design}, 19(3): 291--314.

\bibitem[{Maler, Nickovic, and Pnueli(2007)}]{maler2007synthesizing}
Maler, O.; Nickovic, D.; and Pnueli, A. 2007.
\newblock On synthesizing controllers from bounded-response properties.
\newblock In \emph{International Conference on Computer Aided Verification
  (CAV-07)}, 95--107. Springer.

\bibitem[{Markey(2004)}]{markey04pastfree}
Markey, N. 2004.
\newblock Past is for free: on the complexity of verifying linear temporal
  properties with past.
\newblock \emph{Acta Informatica}, 40(6-7): 431--458.

\bibitem[{Pnueli(1977)}]{pnueli1977temporal}
Pnueli, A. 1977.
\newblock The temporal logic of programs.
\newblock In \emph{{Proceedings of the 18th Annual Symposium on Foundations of
  Computer Science (SFCS-77)}}, 46--57. IEEE.

\bibitem[{Pnueli and Rosner(1989{\natexlab{a}})}]{pnueli1989synthesis}
Pnueli, A.; and Rosner, R. 1989{\natexlab{a}}.
\newblock On the synthesis of an asynchronous reactive module.
\newblock In \emph{Proceedings of the International Colloquium on Automata,
  Languages, and Programming (ICALP-89)}, 652--671. Springer.

\bibitem[{Pnueli and Rosner(1989{\natexlab{b}})}]{PnueliR89}
Pnueli, A.; and Rosner, R. 1989{\natexlab{b}}.
\newblock On the Synthesis of an Asynchronous Reactive Module.
\newblock In Ausiello, G.; Dezani{-}Ciancaglini, M.; and Rocca, S. R.~D., eds.,
  \emph{Proceedings of the 16th International Colloquium on Automata, Languages
  and Programming}, volume 372 of \emph{Lecture Notes in Computer Science},
  652--671. Springer.

\bibitem[{Rosner(1992)}]{rosner1992modular}
Rosner, R. 1992.
\newblock \emph{Modular synthesis of reactive systems}.
\newblock Ph.D. thesis, Weizmann Institute of Science.

\bibitem[{Safra(1988)}]{safra1988complexity}
Safra, S. 1988.
\newblock On the Complexity of omega-Automata.
\newblock In \emph{Proceedings of the 29th Annual Symposium on Foundations of
  Computer Science (FOCS-88)}, 319--327.

\bibitem[{Sistla(1985)}]{sistla1985characterization}
Sistla, A.~P. 1985.
\newblock On characterization of safety and liveness properties in temporal
  logic.
\newblock In \emph{Proceedings of the Fourth Annual {ACM} Symposium on
  Principles of Distributed Computing (PODC-85)}, 39--48.

\bibitem[{Sistla(1994)}]{sistla1994safety}
Sistla, A.~P. 1994.
\newblock Safety, liveness and fairness in temporal logic.
\newblock \emph{Formal Aspects of Computing}, 6(5): 495--511.

\bibitem[{Sistla and Clarke(1985)}]{sistla1985complexity}
Sistla, A.~P.; and Clarke, E.~M. 1985.
\newblock The complexity of propositional linear temporal logics.
\newblock \emph{Journal of the ACM (JACM)}, 32(3): 733--749.

\bibitem[{Thomas(1988)}]{thomas1988safety}
Thomas, W. 1988.
\newblock Safety-and liveness-properties in propositional temporal logic:
  characterizations and decidability.
\newblock \emph{Banach Center Publications}, 1(21): 403--417.

\bibitem[{van Emde~Boas et~al.(1997)}]{van1997convenience}
van Emde~Boas, P.; et~al. 1997.
\newblock The convenience of tilings.
\newblock \emph{Lecture Notes in Pure and Applied Mathematics}, 331--363.

\bibitem[{Vardi and Stockmeyer(1985)}]{VardiS85}
Vardi, M.~Y.; and Stockmeyer, L.~J. 1985.
\newblock Improved Upper and Lower Bounds for Modal Logics of Programs:
  Preliminary Report.
\newblock In Sedgewick, R., ed., \emph{Proceedings of the 17th Annual {ACM}
  Symposium on Theory of Computing}, 240--251. {ACM}.

\bibitem[{Vardi and Wolper(1986)}]{vardi1986automata}
Vardi, M.~Y.; and Wolper, P. 1986.
\newblock An automata-theoretic approach to automatic program verification.
\newblock In \emph{Proceedings of the First Symposium on Logic in Computer
  Science}, 322--331. IEEE Computer Society.

\bibitem[{Vardi and Wolper(1994)}]{vardi1994reasoning}
Vardi, M.~Y.; and Wolper, P. 1994.
\newblock Reasoning about infinite computations.
\newblock \emph{Information and Computation}, 115(1): 1--37.

\bibitem[{Zhu et~al.(2017)Zhu, Tabajara, Li, Pu, and Vardi}]{ZhuTLPV17}
Zhu, S.; Tabajara, L.~M.; Li, J.; Pu, G.; and Vardi, M.~Y. 2017.
\newblock {A Symbolic Approach to Safety LTL Synthesis}.
\newblock In Strichman, O.; and Tzoref{-}Brill, R., eds., \emph{Proceedings of
  the 13th International Haifa Verification Conference}, 147--162. Springer.

\end{thebibliography}

\ifappendix
  \clearpage
  \appendix

\section{Tiling Problems}

We define the notions of tiling structure, infinite corridor tiling and
tiling games.

\begin{definition}[Tiling Structure]
\label{def:tilingstruct}
  A \emph{tiling structure} is a tuple $\tiling = \seq{ T, t_{\bor}, H,$
  $V}$, where $T$ is a finite set of elements, called \emph{tiles},
  $t_{\bor} \in T$ is the \emph{border tile}, and $H , V \subseteq T \times
  T$ are the \emph{horizontal} and the \emph{vertical relations},
  respectively.
\end{definition}
\begin{definition}[Infinite Corridor Tiling]
\label{def:rectangletiling}
  Let $\tiling = \seq{ T, t_{\bor}, H, V}$ be a tiling structure, and let
  $n\in \N$. We define an \emph{infinite $n$-corridor tiling} for $\tiling$
  as a function $f : \N \times [0,n) \to T$ that associates a tile in $T$
  with every position of the infinite corridor of height $n$ in such a way
  that:
  \begin{enumerate}
    \item 
      the horizontal relation is satisfied:
      \begin{align*}
        \forall x \in [0,k-1) \ \forall y \in [0,n) \suchdot
        f(x,y) H f(x+1,y);
      \end{align*}
    \item 
      the vertical relation is satisfied:
      \begin{align*}
        \forall x \in [0,k) \ \forall y \in [0,n-1) \suchdot
        f(x,y) V f(x,y+1);
      \end{align*}
    \item 
      the top and bottom borders of the $n \times k$-rectangle are tiled with
      $t_{\bor}$:
      \begin{align*}
        \forall x \in [0,k) \suchdot f(x,0) = f(x,n-1) = t_{\bor}
      \end{align*}
  \end{enumerate}
\end{definition}

Tiling games~\cite{chlebus1986domino} consider two players:
\emph{Constructor}, whose goal is to build a tiling for the tiling
structure $\tiling$, and \emph{Saboteur}, trying to prevent this from
happening.
The two players play one at a time (with Constructor being the first
one to play), choose a tile from $T$ and position it on the
tiling structure $\tiling$ in a precise order: the first position is the
one at coordinates $(0,0)$, the second position is the one at coordinates
$(0,1)$, and so on and so forth. When a column is entirely tiled, the game
proceeds on the next column.
If there is no tile fitting the next position or the definition of tiling
(\cref{def:rectangletiling}) is violated, then Saboteur wins.
Otherwise, Constructor wins. 

Given an $n\in\N$ encoded in unary (resp. in binary) and a tiling structure
$\tiling$, we call \textnormal{INFCORR-GAME} (resp.
\textnormal{EXP-INFCORR-GAME}) the problem of finding whether there exists
a strategy for Constructor for building an infinite $n$-corridor tiling for
$\tiling$.

\begin{proposition}[\cite{chlebus1986domino}]
\label{prop:tilinggames}
  It holds that:
  \begin{itemize}
    \item \textnormal{INFCORR-GAME} is \EXPTIME-complete.
    \item \textnormal{EXP-INFCORR-GAME} is \EXPTIME[2]-complete.
  \end{itemize}
\end{proposition}

\section{Proofs}

In this section, for any $n\in\N$, we inductively define the formula
$\ltl{X}^n\phi$ as follows: $\ltl{X}^0\phi \coloneqq \phi$ and $\ltl{X}^{n+1}\phi
\coloneqq \ltl{X} (\ltl{X}^n \phi)$. We also define $\ltl{Y}^n\phi$ and
$\ltl{wY}^n\phi$ in a similar way.

\subsection{General Theorem on Finite Traces}

\thfininf*
\begin{proof}
  By definition of suffix independence,
  $\lang(\phi)=\lang(\phi)_F\cdot(2^\Sigma)^\omega$. Therefore, if
  $\lang(\phi)_F=\emptyset$, then $\lang(\phi)=\emptyset$. Otherwise, if
  $\lang(\phi)_F\neq\emptyset$, then $\lang(\phi)\neq\emptyset$.
\end{proof}

\thsmpsafetyfin*
\begin{proof}
  We prove the opposite direction, that is, if $\seq{\sigma_0}\not\in\lang$
  for all $\seq{\sigma_0}\in(2^\Sigma)^+$ of length $1$, then
  $\lang=\emptyset$.
  
  Since $\lang$ is by hypothesis a safety language, by
  \cref{def:cosafelang,def:safelang}, it holds that, for all
  $\sigma\in(2^\Sigma)^+$, if $\sigma\not\in\lang$ then there exists an
  $i<|\sigma|$ such that $\sigma_{[0,i]}\cdot\sigma'\not\in\lang$, for all
  $\sigma'\in(2^\Sigma)^+$.  By hypothesis, we have that
  $\seq{\sigma_0}\not\in\lang$ for all $\seq{\sigma_0}\in(2^\Sigma)^+$ of
  length $1$. Therefore, it holds that:
  \begin{align*}
    &\forall \sigma\in(2^\Sigma)^+ \suchdot (
      |\sigma|=1 \to \\
      &\qquad\exists i<|\sigma| \suchdot \forall
      \sigma'\in(2^\Sigma)^+\suchdot \sigma_{[0,i]}\cdot\sigma'
      \not\in\lang
    ) \\
    \Iff \ 
    &\forall \sigma\in(2^\Sigma)^+ \suchdot (
      |\sigma|=1 \to \forall \sigma'\in(2^\Sigma)^+\suchdot
      \sigma_{[0,0]}\cdot\sigma' \not\in\lang
    )
  \end{align*}
  But this is equivalent to say that $\forall \sigma\in(2^\Sigma)^+
  \suchdot ( \sigma\not\in\lang)$. That is, $\lang=\emptyset$.
\end{proof}

\thsmpsafetyfinlogic*
\begin{proof}
  In the current proof, we consider only finite trace semantics.  We first
  prove the case for \safetyltl: the case for \LTLwxg follows since it is
  a syntactical fragment of \safetyltl. We proceed by induction on the
  structure of $\phi$:
  \begin{itemize}
    \item if $\phi = p$ (resp. $\phi = \lnot p$), with $p \in
      \Sigma$, then by hypothesis $p \in \sigma_0$ (resp. $p \not\in
      \sigma_0$), and thus $\seq{\sigma_0} \models p$ (resp.
      $\seq{\sigma_0} \models p$);
    \item if $\phi = \phi_1 \land \phi_2$, then by hypothesis $\sigma
      \models \phi_1$ and $\sigma \models \phi_2$. By inductive hypothesis,
      $\seq{\sigma_0} \models \phi_1$ and $\seq{\sigma_0} \models \phi_2$,
      that is $\seq{\sigma_0} \models \phi_1 \land \phi_2$;
    \item if $\phi = \phi_1 \lor \phi_2$, the proof is the same for
      the case of conjunctions;
    \item if $\phi = \ltl{wX \phi_1}$, then by the semantics of the
      $\ltl{wX}$ operator, it holds that $\seq{\sigma_0} \models \ltl{wX
      \phi_1}$;
    \item $\phi = \ltl{\phi_1 R \phi_2}$, then by the semantics of the
      $\ltl{R}$ operator, it holds that either $\sigma,i \models \phi_2$
      for all $0\le i < |\sigma|$ or there exists a $0 \le j < |\sigma|$
      such that $\sigma,j \models \phi_1$ and $\sigma,k \models \phi_2$ for
      all $0 \le k \le j$.  We divide in cases:
      \begin{itemize}
        \item in the first case, it holds that $\seq{\sigma_0} \models
          \phi_2$; thus for all $0 \le i < |\seq{\sigma_0}|$ it holds that
          $\seq{\sigma_0},i \models \phi_2$, that is $\seq{\sigma_0}
          \models \ltl{\phi_1 R \phi_2}$;
        \item in the second case, it holds that $\seq{\sigma_0} \models
          \phi_1 \land \phi_2$; thus there exists a $0 \le
          j < |\seq{\sigma_0}|$ such that $\seq{\sigma_0},j \models \phi_1$
          and $\seq{\sigma_0},k \models \phi_2$ for all $0\le k \le j$;
          that is $\seq{\sigma_0} \models \ltl{\phi_1 R \phi_2}$.
      \end{itemize}
  \end{itemize}

  We now prove the same for \Galpha. Let $\phi\in\Galpha$.  Suppose that
  $\sigma\models\phi$ for some $\sigma \in (2^\Sigma)^+$. Then, by the
  semantics of the $\ltl{G}$ operator over finite traces, it holds that
  $\sigma,i \models \alpha$ for all $0 \le i < |\sigma|$. Since $\alpha$ is
  a formula of pure past \LTLP, this means (for $i=0$) that $\seq{\sigma_0}
  \models \alpha$. Since $|\seq{\sigma_0}|=1$, this means that
  $\seq{\sigma_0},i \models \alpha$ for all $0\le i < |\seq{\sigma_0}|$. By
  the semantics of the $\ltl{G}$ operator on finite traces, we have that
  $\seq{\sigma_0} \models \ltl{G}(\alpha)$.
\end{proof}

\subsection{Satisfiability and Validity}

\lemmasatendt*
\begin{proof}
\nb{M: changed, to check}
  We first prove the
  left-to-right direction.
  Suppose that $\phi$ is
  satisfiable over finite traces, that is, there exists a finite
  trace $\sigma\in(2^\Sigma)^+$ such that $\sigma\models\phi$. Let $m=|\sigma|$. Moreover, set $\Sigma'
  \coloneqq \Sigma \cup \set{\endt}$. We define
  $\sigma'\in(2^{\Sigma'})^\omega$ to be any infinite trace such
  that:
  \begin{enumerate}
    \item $\sigma'_i = \sigma_i$,
    for every $0\le i < m$; and
    \item $\endt \in \sigma'_m$.
  \end{enumerate}
  It follows that $\sigma' \models \lnot\endt \land (\lnot
  \endt)\ltl{U}(\endt)$.
To prove that
  $\sigma' \models f(\phi)$, we show by induction on the structure of $\phi \in \LTL$ that $\sigma,i \models \phi$ iff $\sigma',i \models f(\phi)$, for every $0\le i< m$.
\begin{itemize}
    \item Let $\phi = p$, for any $p\in\Sigma$. We have that $\sigma,i \models p$ iff $p \in \sigma_i$. Since by definition $\sigma'_i = \sigma_i$, the previous step is equivalent to $p\in\sigma'_i$. Given that $f(p)=p$, this means $\sigma', i\models f(p)$.
    \item Let $\phi = \lnot p$, for any $p\in\Sigma$. The proof is the same as in the previous case.
    \item Let $\phi = \phi_1 \land \phi_2$. We have $\sigma,i \models \phi_1 \land \phi_2$ iff $\sigma,i \models \phi_1$ and $\sigma,i \models \phi_2$. Equivalently, by inductive hypothesis, $\sigma',i\models f(\phi_1)$ and $\sigma',i\models f(\phi_2)$, \ie $\sigma',i\models f(\phi_1) \land f(\phi_2)$.
      By definition, this means $\sigma',i\models f(\phi_1 \land \phi_2)$.
    \item Let $\phi = \phi_1 \lor \phi_2$. The proof is similar to
      the previous case.
    \item Let $\phi = \ltl{X} \phi_1$.
    We have that $\sigma,i \models \ltl{X \phi_1}$ iff $i+1 < m$ and $\sigma,i+1 \models
      \phi_1$.
      By inductive hypothesis, this is equivalent to $i+1 < m$ and $\sigma',i+1 \models f(\phi_1)$.
      By definition of $\sigma'$, the previous step means that $\sigma',i+1 \models \lnot \endt$ and $\sigma',i+1 \models f(\phi_1)$, \ie $\sigma',i+1 \models \lnot\endt \land f(\phi_1)$. Equivalently, $\sigma',i \models \ltl{X(! \endt && f(\phi_1))}$, that is, $\sigma',i \models \ltl{f(X \phi_1)}$.
    \item Let $\phi = \ltl{wX}\phi_1$.
    
    $(\Rightarrow)$
    Suppose that $\sigma,i \models \ltl{wX}\phi_1$, \ie either $i+1 = m$ or $\sigma,i+1 \models \phi_1$.
      We reason by cases. If $i+1=m$, then by definition of $\sigma'$ we have that $\sigma',i+1 \models \endt$. If instead $i+1<m$ and $\sigma,i+1 \models \phi_1$, then by inductive hypothesis we have that $\sigma',i+1 \models f(\phi_1)$. Therefore, it holds that $\sigma',i+1 \models \endt \lor f(\phi_1)$, that is, $\sigma',i \models \ltl{X(! \endt -> f(\phi_1))}$.
      
    $(\Leftarrow)$
     Suppose that $\sigma',i \models f(\ltl{wX}(\phi_1))$, \ie $\sigma',i \models \ltl{X(! \endt -> f(\phi_1))}$, meaning that $\sigma',i+1 \models \ltl{! \endt -> f(\phi_1)}$. Equivalently, $\sigma',i+1 \models \ltl{\endt \lor f(\phi_1)}$.
     We reason by cases. If $\sigma',i+1 \models \ltl{\endt}$, then we have by definition of $\sigma'$ that $i + 1 = m$.
     If $\sigma', i+1 \models \ltl{f(\phi_1)}$, then either $i + 1 = m$ or, by inductive hypothesis, $\sigma, i+1 \models \phi_1$. In either case, we obtain $\sigma, i \models \ltl{wX}(\phi_1)$.
    \item
    Let $\phi = \ltl{\phi_1 U \phi_2}$.
    
    $(\Rightarrow)$
    If $\sigma,i
      \models \ltl{\phi_1 U \phi_2}$, then there exists $i\le j<m$ such that $\sigma,j \models \phi_2$ and $\sigma,k\models \phi_1$, for every $i\le k<j$.
      By
      inductive hypothesis, we have $\sigma',j\models f(\phi_2)$ and
      $\sigma',k\models f(\phi_1)$, for all $i\le k<j$.
      Since
      $j<m$, by definition of $\sigma'$ we have that $\sigma', n \models
      \lnot \endt$, for every $i \le n \leq j$. Therefore, $\sigma',i\models\ltl{(! \endt && f(\phi_1)) U (! \endt
      && f(\phi_2))}$.
      
      $(\Leftarrow)$
      Suppose that $\sigma', i \models \ltl{f (\phi_1 U \phi_2)}$, \ie $\sigma',i\models\ltl{(! \endt && f(\phi_1)) U (! \endt
      && f(\phi_2))}$. Then, there exists $i \leq j$ such that $\sigma', j \models  \ltl{f(\phi_2)}$ and $\sigma', k \models \ltl{f(\phi_1)}$, for all $i \leq k < j$, with $\sigma', n \models \ltl{! \endt}$, for every $i \leq n \leq j$.
      By construction of $\sigma'$, the previous step implies that $j < m$. Thus, by inductive hypothesis we have that there exists $i \leq j < m$ such that $\sigma, j \models  \ltl{\phi_2}$ and $\sigma, k \models \phi_1$, for all $i \leq k < j$, that is, $\sigma, i \models  \ltl{\phi_1 U \phi_2}$.
      \item
      Let $\phi = \ltl{\phi_1 R \phi_2}$.
      
      $(\Rightarrow)$
      If $\sigma, i \models \ltl{ \phi_1 R \phi_2}$, then
      either $\sigma, j \models \phi_2$,
      for all $i \le j < m$, or there exists $i \le k < m$
      such that $\sigma, k \models \phi_1$ and $\sigma, j \models \phi_2$, for all $i \le j \le k$.
      By inductive hypothesis, this implies that:
      either $(a)$ $\sigma', j \models f(\phi_2)$,
      for all $i \le j < m$,
      or
      $(b)$
      there exists a $i \le k < m$
      such that $\sigma, k \models f(\phi_1)$ and $\sigma, j \models f(\phi_2)$ for
      all $i \le j \le k$.
      We consider the two cases.
      \begin{itemize}
        \item[$(a)$] Since, by construction, it holds that $\sigma',m
      \models \endt$ and $\sigma', j \models \lnot \endt$, for every $0 \leq j < m$,
      we
      have $\sigma',i \models \ltl{(! \endt && f(\phi_2) ) U (! \endt && f(\phi_2) && X(\endt))}$.
      \item[$(b)$] It holds that $\sigma', k \models
          f(\phi_1) \land f(\phi_2)$, for some $i \leq k < m$, and $\sigma, j \models f(\phi_2)$ for
      all $i \le j \le k$. Moreover, by construction, $\sigma', j \models \lnot \endt$, for all $i \le j \le k$. Hence, $\sigma', i \models \ltl{(! \endt && f(\phi_2) ) U (! \endt && f(\phi_1) \land f(\phi_2))}$.
      \end{itemize}
     In either case, we obtain $\sigma', i \models f(\ltl{\phi_1 R \phi_2})$.
     
     $(\Leftarrow)$
     Suppose that $\sigma', i \models f(\ltl{\phi_1 R \phi_2})$, \ie $\sigma', i \models
     \ltl{(! \endt && f(\phi_2) ) U (! \endt && f(\phi_2) && X(\endt))}
     \lor
     \ltl{(! \endt && f(\phi_2) ) U (! \endt && f(\phi_1) && f(\phi_2))}$. This means the following:
     either
     $(a)$
 there exists $i \leq j$ such that 
 $\sigma', j \models \lnot \endt$,
 $\sigma', j \models f(\phi_2)$ and $\sigma', j+1 \models \endt$, where in addition $\sigma', j' \models \lnot \endt$ and $\sigma', j' \models f(\phi_2)$, for every $i \leq j' < j$;
      or
      $(b)$ 
      there exists $i \leq k$
      such that
       $\sigma', k \models \lnot \endt$,
       $\sigma', k \models f(\phi_1)$ and $\sigma, k \models f(\phi_2)$, where also $\sigma', k' \models \lnot \endt$ and $\sigma', k' \models f(\phi_2)$, for every $i \le k' < k$.
       By construction of $\sigma'$, the previous step implies that $j +  1 = m$ and $k < m$.
We now distinguish the two cases.
       \begin{itemize}
       	\item[$(a)$] We have in particular that there exists $i \leq j < m$, with $j + 1 = m$, such that $\sigma', j \models f(\phi_2)$ and $\sigma', j' \models f(\phi_2)$, for every $i \leq j' < j$. Hence $\sigma', j \models f(\phi_2)$, for every $i \leq j < m$. By inductive hypothesis, this implies $\sigma, j \models \phi_2$, for every $i \leq j < m$.
		\item[$(b)$]
	We have in particular that there exists $i \leq k < m$
      such that
       $\sigma', k \models f(\phi_1)$ and $\sigma', k' \models f(\phi_2)$, for every $i \le k' \leq k$. Hence,
       by inductive hypothesis,
       there exists $i \leq k < m$
      such that
       $\sigma, k \models \phi_1$ and $\sigma', k' \models \phi_2$, for every $i \le k' \leq k$.
       \end{itemize} 
    In either case, we obtain that $\sigma, i \models \ltl{\phi_1 R \phi_2}$.
  \end{itemize}
Therefore, $\sigma' \models f(\phi)$, and hence $\sigma' \models g(\phi)$.

  We now prove the right-to-left direction. Suppose that $g(\phi)$ is satisfiable over infinite traces, \ie there exists an infinite trace
  $\sigma'\in(2^{\Sigma'})^\omega$ such that $\sigma'\models g(\phi)$.
  Since by definition $g(\phi)\coloneqq\ltl{! \endt && f(\phi) && (! \endt)
  U (\endt)}$, it holds that there exists an $m\in\N$ such that $\sigma',m
  \models\endt$ and $\sigma',i\models\lnot\endt$ for each $0\le i< m$.  We
  define the finite trace $\sigma\in(2^\Sigma)^+$ of length $m$ as follows:
  $\sigma_i = \sigma'_i \cap \Sigma$, for any $0\le i<m$.
  Similarly to the converse direction above, it can be shown by induction on the structure of $\phi \in \LTL$ that $\sigma', i \models f(\phi)$ iff $\sigma, i \models \phi$, for every $0 \leq i < m$.
  Thus, we obtain
  $\sigma \models \phi$.
\end{proof}

\lemmasatfalpha*
\begin{proof}
  \emph{(Membership)} It follows from the fact that $\Falpha$ is
  a syntactical fragment of \LTLP and that satisfiability of \LTLP is
  \PSPACE-complete~\cite{sistla1985complexity}.

  \emph{(Hardness)}
  We reduce the satisfiability problem (over finite traces) of \LTLFP to
  the satisfiability problem (over infinite traces) of \Falpha. For any
  $\phi\in\LTLFP$ it holds that $\phi$ is satisfiable iff $\ltl{F(\phi)}$
  is satisfiable over infinite traces. Since \LTLFP satisfiability is
  \PSPACE-complete, it follows that \Falpha satisfiability (over infinite
  traces) is \PSPACE-hard.
\end{proof}

\valinffalpha*
\begin{proof}
  First of all, we note that the validity problem of \LTL over finite
  traces is \PSPACE-complete. This follows from the fact that, for each
  formula $\phi\in\LTL$, $\phi$ is not valid over finite traces iff
  $\lnot\phi$ is satisfiable over finite traces. Since \LTL is closed under
  complement, we can solve validity with an algorithm for satisfiability,
  and \viceversa. Therefore, since satisfiability of \LTL over finite
  traces is \PSPACE-complete~\cite{DeGiacomoV13}, the validity of \LTL over
  finite traces is \PSPACE-complete as well. By observing that $\phi$ is
  valid iff $\phi'$ is valid (where $\phi'$ is the formula obtained from
  $\phi$ by replacing each $\ltl{X}$ and $\ltl{U}$ operator with $\ltl{Y}$
  and $\ltl{S}$) it follows the \PSPACE-completeness (over finite traces)
  of \LTLFP as well.

  We now prove the \PSPACE-completeness for the validity problem of \Falpha
  over infinite traces.

  \emph{(membership)} It follows from the \PSPACE-completeness of \LTLP
  validity over finite traces.

  \emph{(hardness)} We reduce the validity problem of \LTLFP (over finite
  traces) to validity over finite traced of \Falpha. Let $\phi$ be any
  formula of \LTLFP. We prove that $\phi$ is valid over finite traces iff
  $\ltl{F}(\phi)$ is valid over infinite traces. We have that:
  \begin{align*}
     &\phi \mbox{ is valid over finite trace} \\
    \Iff \ 
     &\forall\sigma\in(2^\Sigma)^+ \suchdot \sigma\models\phi \\
    \Iff \ 
     &\forall\sigma\in(2^\Sigma)^+ \suchdot \forall\sigma\in(2^\Sigma)^\omega
     \suchdot (\sigma\cdot\sigma',|\sigma| \models\phi) \\
    \Iff \
     &\forall\sigma\in(2^\Sigma)^+ \suchdot \forall\sigma\in(2^\Sigma)^\omega
     \suchdot (\sigma\cdot\sigma' \models\ltl{F}\phi) \\
    \Iff \
     &\forall\sigma\in(2^\Sigma)^\omega \suchdot \sigma \models \ltl{F}\phi \\
    \Iff \
     &\ltl{F}\phi \mbox{ is valid over infinite trace}
  \end{align*}
\end{proof}

\lemfininflogics*
\begin{proof}
  From now on, with $\lang(\phi)$ (resp. $\lang(\phi)_F$) we denote the
  language of $\phi$ over infinite (resp. finite) traces.  

  We start with the \Falpha fragment. We first prove the inclusion
  $\lang(\phi)\subseteq\lang(\phi)_F\cdot(2^\Sigma)^\omega$.  Let $\phi$ be
  a formula of type $\ltl{F}(\alpha)$ with $\alpha\in\LTLFP$, and let
  $\sigma$ be a model of $\phi$. By the semantics of the $\ltl{F}$
  operator, and since $\alpha$ is a pure past formula of \LTLP, it holds
  that there exists a $k\ge 0$ such that $\sigma_{[0,k]} \models \phi$,
  where $\sigma_{[0,k]}$ is the prefix from $0$ to $k$ of $\sigma$. This is
  equivalent to say that there exists a finite trace
  $\sigma'\in(2^\Sigma)^+$ such that $\sigma'\models\phi$ and
  $\sigma'\cdot\sigma'' = \sigma$ for any $\sigma''\in(2^\Sigma)^\omega$.
  Therefore $\sigma\in\lang(\phi)_F\cdot(2^\Sigma)^\omega$.  We now prove
  the inclusion $\lang(\phi)_F\cdot(2^\Sigma)^\omega \subseteq
  \lang(\phi)$. Let $\sigma$ be a trace in
  $\lang(\phi)_F\cdot(2^\Sigma)^\omega$. By definition of $\sigma$, it
  holds that $\sigma=\sigma'\cdot\sigma''$ where $\sigma'\in(2^\Sigma)^+$
  is such that $\sigma'\models\ltl{F(\alpha)}$ and
  $\sigma''\in(2^\Sigma)^\omega$.  By the semantics of the $\ltl{F}$
  operator, it holds that there exists a $k\ge 0$ such that
  $\sigma',k\models\alpha$. Since $\alpha$ is a pure past formula of \LTLP,
  this is equivalent to say that $\sigma,k \models\alpha$, that is
  $\sigma\models\ltl{F(\alpha)}$, and thus $\sigma\in\lang(\phi)$.

  The case for \cosafetyltl is proved by Lemma $1$
  in~\cite{DBLP:conf/fossacs/CimattiGGMT22} (cf. also \cite[Lemma
  4.11]{artale22TOCLarxivsubmission}).
\end{proof}

\corsmpsafety*
\begin{proof}
  \emph{(membership)} Let
  $\mathbb{L} \in \{ \safetyltl, \Galpha$, $\LTLwxg \}$ and
  $\phi\in\mathbb{L}$.
  It
  suffices to guess an assignment for the initial state of a candidate
  trace and check if it satisfies $\phi$. If such an assignment is found,
  then it means that the formula is satisfiable, otherwise, since $\phi$ is
  a \emph{safety} formula, by \cref{th:smp:safety:fin}, it is
  unsatisfiable.
  
  \emph{(hardness)} It simply follows from a reduction of the SAT problem.
\end{proof}

\subsection{Realizability}

\lemmarealendt*
\begin{proof}
  We first prove the left-to-right direction. Suppose that $\phi$ is
  realizable over finite traces. Then there exists a strategy $s
  : (2^\Uset)^+ \to 2^\Cset$ such that, for any infinite sequence
  $\Uncontr=\seq{\Uncontr_0, \Uncontr_1, \dots}$ in $(2^\Uset)^\omega$, it
  holds that there exists a $k\in\N$ such that the prefix of
  $\res(s,\Uncontr)$ from $0$ to $k$ is a model of $\phi$. For any
  $\overline{\Uncontr}=\seq{\Uncontr_0,\Uncontr_1,\dots,\Uncontr_n}$, we
  define the strategy $s' : (2^{\Uset'})^+ \to (2^{\Cset'})$ as follows:
  \begin{align*}
    s'(\overline{\Uncontr}) =
    \begin{cases}
      \set{\endt} & \mbox{if } s'(\seq{\Uncontr_0,\dots,\Uncontr_k}) \models \phi \\
      & \mbox{for some } k\le n  \\
      s(\overline{\Uncontr}) & \mbox{otherwise}
    \end{cases}
  \end{align*}
  Notice that, in the second case of the definition of $s'$, the
  proposition letter $\endt$ is absent from
  $s(\seq{\Uncontr_0,\dots,\Uncontr_n})$ and thus it is supposed to be
  false in $s'(\seq{\Uncontr_0,\dots,\Uncontr_n})$.
  It is simple to see that, for any infinite sequence
  $\Uncontr=\seq{\Uncontr_0, \Uncontr_1, \dots}$ in $(2^\Uset)^\omega$,
  there exists a $k\in\N$ such that the prefix of $\res(s',\Uncontr)$ from
  $0$ to $k$ is a model of $g(\phi)$.

  We now prove the right-to-left direction. Suppose that $g(\phi)$ is
  realizable over infinite traces. There exists a strategy $s'
  : (2^{\Uset'})^+ \to (2^{\Cset'})$ such that $\res(s',\Uncontr) \models
  \phi$, for any infinite sequence $\Uncontr$. Since by definition $g(\phi)
  \coloneqq \ltl{! \endt && f(\phi) && (! \endt) U (\endt)}$, it holds
  that, for any infinite sequence $\Uncontr$, there exists a $k\in\N$ such
  that the prefix from $0$ to $k$ of $\res(s',\Uncontr)$ is a model of
  $f(\phi)$. By the induction proof of \cref{lemma:sat:endt}, it holds that
  the projection of this prefix into the variables in $\Uset \cup \Cset$ is
  a model of $\phi$. Therefore, the strategy $s : (2^{\Uset})^+
  \to(2^{\Cset})$ defined as the projection of $s'$ to variables in
  $\Uset\cup\Cset$ is such that, for any infinite sequence
  $\Uncontr=\seq{\Uncontr_0, \Uncontr_1, \dots}$ in $(2^\Uset)^\omega$,
  there exists a $k\in\N$ such that the prefix of $\res(s,\Uncontr)$ from
  $0$ to $k$ is a model of $\phi$.
\end{proof}

\realcosafetyltl*
\begin{proof}
  \emph{(Membership)}
  It follows from the \EXPTIME[2]-membership of \LTL realizability on infinite traces~\cite{pnueli1989synthesis}.

  \emph{(Hardness)}
  From \cref{lemma:real:endt} and the fact that the size of $g(\phi)$ is
  polynomial in the size of $\phi$, it follows that we can reduce the
  realizability problem of \LTL over finite traces (which is
  \EXPTIME[2]-complete~\cite{DeGiacomoV15}) to the realizability of
  \cosafetyltl over infinite traces.
\end{proof}

\realgalpha*
\begin{proof}
  \emph{(Membership)} It follows from the fact that \Galpha is
  a syntactical fragment of \LTLEBRP with no
  constants~\cite{cimatti2021extended} and that the realizability problem
  of this fragment is \EXPTIME-complete~\cite{cimatti2021extended}.

  \emph{(Hardness)}  
  We reduce \textnormal{INFCORR-GAME} to \Galpha satisfiability on infinite
  traces.  For a tiling structure $\tiling$ and $n \in \mathbb{N}$, given
  in unary, we will define a \Galpha formula $\phi_{n,\tiling}$ such that:
  $(i)$ $\phi_{n,\tiling}$ has length polynomial in $n$ and in the size of
  $\tiling$; $(ii)$ $\phi_{n,\tiling}$ is realizable on infinite traces iff
  Constructor can force an infinite $n$-corridor tiling for $\tiling$.

  We will make the natural correspondence between Controller and Environment
  players (of realizability) with Constructor and Saboteur players of tiling
  games.
  Finally, for proving the \EXPTIME-hardness, we will use the variant with
  $n$ encoded in unary.
  
  As noted in~\cite{cimatti2021extended}, there are three main problems
  that have to be addressed when dealing with a reduction from tiling games
  into realizability problems:
  \begin{enumerate}
    \item the variables under the control of Environment and Controller
      players are disjoint while Saboteur and Constructor choose tiles from
      the same set $T$;
    \item a round between Environment and Controller creates a state of
      a trace (which, in our reduction, corresponds to a cell of a tiling),
      while a round between Saboteur and Constructor constists of two cells
      of a tiling;
    \item Environment moves before Controller, while Constructor moves before
      Saboteur (this point as well as the previous ones were already noted in
      \cite{cimatti2021extended}).
  \end{enumerate}

  We will define $\phi_{n,\tiling}^{\game}$ over the alphabet $\Sigma
  = \Uset\cup\Cset$, where $\Uset \coloneqq \set{t^u \mid t \in T}$ and
  $\Cset \coloneqq \set{t^c \mid t \in T} \cup \set{b}$. This solves the
  first problem. Note that the proposition letter $b$ is set to be
  \emph{controllable}: this because we want to ensure that, whenever the
  formula is unrealizable, the reason is not a wrong marking of the top and
  bottom border but, rather, the nonexistence of a strategy for Constructor
  for building a tiling.
  
  For solving the second problem, we have to impose that, for any time point
  $i$, Environment player plays at round $i$ iff Controller player plays at
  round $i+1$ and they never play at the same round.  

  As for the third problem, it suffices to force the formula $\firstplayer
  \coloneqq \bigvee_{t\in T}t^c$ to be true at the initial time point.
  
  We define the following formulas:
  \begin{itemize}
    \item \emph{Saboteur} plays at round $t$ iff \emph{Constructor} plays at
      round $t\pm 1$, and the two player never play in the same round:
      \begin{align*}
        \altern
          &\coloneqq
          \bigvee_{t\in T} (t^u \iff \ltl{Y}(\bigvee_{t\in T} t^c)) \land
          \bigvee_{t\in T} (t^u \iff \bigwedge_{t\in T} \lnot t^c)
      \end{align*}
    \item The current position corresponds to a cell of the first column:
      \begin{align*}
        \incolumn1
          &\coloneqq
          \ltl{wY^{n} \false}
      \end{align*}
    \item The current cell corresponds to the bottom-left angle of the
      infinite corridor:
			\begin{align*}
        \bottomleftangle
          &\coloneqq
          \ltl{wY \bot}
			\end{align*}
    \item The current cell corresponds to the top-left angle of the
      infinite corridor:
			\begin{align*}
        \topleftangle
          &\coloneqq
          \ltl{wY^n \bot && Y^{n-1}\top}
			\end{align*}
    \item A given cell is tiled with exactly one tile:
      \begin{equation*}
        \tile
        \coloneqq
        (\bigvee_{t \in T} t \land \bigwedge_{\substack{t,t' \in T \\ t \neq
        t'}} (\lnot t \lor \lnot t') )
	  	\end{equation*}
    \item The first and the last position of the current column are marked
      with $b$ while all the positions in between are marked with $\lnot b$:
	  		\begin{equation*}
          \bcolumn
          \coloneqq 
          b \land \ltl{Y}^{n-1}(b) \land \bigwedge_{i \in [1, n-2]}
          \ltl{Y}^{i} (\lnot b)
	  		\end{equation*}
     \item The current position is marked with $b$ iff the position to its
       left is marked with $b$, and if the current position is marked with
       $b$ then it is tiled with $t_{\bor}$:
	  		\begin{align*}
          \bordertiling
            &\coloneqq
              \ltl{(b \iff \ltl{Y}^n b)}
              \land
              (b \to t_{\bor}) \\
            &\coloneqq
              \ltl{((! b && \ltl{Y}^n ! b) ||
                   (b && \ltl{Y}^n b)) &&
                   (! b || t_{\bor})}
			\end{align*}
	  \item Tiles respect the horizontal relation:
	  		\begin{equation*}
          \hor
            \coloneqq 
            \bigvee_{(t,t') \in H} (t \land \ltl{Y}^{n} t' ).
	  		\end{equation*}
      \item A tile either stands in a border or it respects the vertical
        relation (the formula $(\ltl{Y} b \land \ltl{Y Y} \lnot b)$ excludes the
        pair consisting of positions $(i,n-1)$ and $(i-1,0)$, which do not
        need to fulfill the vertical relation):
	  			\begin{equation*}
            \ver
              \coloneqq 
              \ltl{(\ltl{Y} b & \ltl{Y Y} ! b)} \lor \bigvee_{(t,t') \in V} (t \land
              \ltl{Y} t' ).
				\end{equation*}
  \end{itemize}
  Finally, we define $\invariant$ as the following formula 
  \begin{align*}
    \invariant \coloneqq \tile \land \bordertiling \land \hor \land \ver
  \end{align*}
  We now give the formula $\phi_{n, \tiling}$ in \Galpha such that
  $\phi_{n, \tiling}$ is realizable iff Constructor has a winning strategy
  in $\tiling$. It is built starting from the following formulas:
  \begin{itemize}
    \item Constructor chooses the tile for the cell corresponding to the
      bottom-left angle:
      \begin{equation*}
        \bottomleftangle \to \firstplayer
      \end{equation*}
    \item for each cell except for the one corresponding to the bottom-left
      angle, at most one player plays in that cell and the is tiled with
      exactly one tile:
      \begin{equation*}
        \lnot\bottomleftangle \to (\altern \land \tile)
      \end{equation*}
    \item the first and the last cell of the first column satisfy $b$ while
      all the other satisfy $\lnot b$:
      \begin{equation*}
        \topleftangle \to \bcolumn
      \end{equation*}
    \item the first column satisfy the vertical relation and each cell
      marked with $b$ is tiled with $t_{\bor}$:
      \begin{equation*}
        \incolumn1 \to (\ver \land (b \to t_{\bor}))
      \end{equation*}
    \item if a cell does not belong to the first column, then
      \bordertiling, the horizontal and the vertical constraints must hold:
      \begin{equation*}
        \lnot \incolumn1 \to \bordertiling \land \hor \land \ver
      \end{equation*}
  \end{itemize}
  We set:
  \begin{align*}
    \phi_{n, \tiling} \coloneqq 
      \ltl{G} \big(  
        &(\bottomleftangle \to \firstplayer) \land {} \\
        &(\lnot\bottomleftangle \to (\altern \land \tile)) \land {} \\
        &(\topleftangle \to \bcolumn) \land {} \\
        &(\incolumn1 \to (\ver \land (b \to t_{\bor}))) \land {} \\
        &(\lnot \incolumn1 \to \bordertiling \land \hor \land \ver)
    \big)
  \end{align*}
  The length of $\phi_{n,\tiling}$ is polynomial in $n$ and $|T|$, and it
  holds that $\phi_{n,\tiling}$ is realizable on infinite traces iff
  Controller has a strategy for forcing an infinite $n$-corridor tiling for
  $\tiling$.
  By \cref{prop:tilinggames}, realizability over infinite traces of the
  \Galpha fragment is \EXPTIME-hard.
\end{proof}

\realduality*
\begin{proof}
  We first prove the case for $\mathbb{L}=\cosafetyltl$ and
  $\mathbb{L}'=\safetyltl$. Let $\phi$ be any formula in \cosafetyltl. From
  now until the end of the proof we assume infinite trace semantics. It
  holds that:
  \begin{align}
  \begin{split}
  \label{eq:proof:dual}
    &\phi \mbox{ is realizable } \\
      \Iff \
    &\exists s : (2^\Uset)^+ \to 2^\Cset \suchdot \forall
    \Uncontr \in (2^\Uset)^\omega \suchdot
    \res(s,\Uncontr) \models \phi \\ \Iff \
    &\lnot\exists s' : (2^\Cset)^* \to 2^\Uset \suchdot \forall
    \Contr \in (2^\Cset)^\omega \suchdot \res(s',\Contr)
    \models \lnot\phi
  \end{split}
  \end{align}
  where $\res(s',\Contr)$ is defined as the sequence
  $\seq{(s'(\epsilon) \cup \Contr_0), (s'(\seq{\Contr_0}) \cup \Contr_1),
  (s'(\seq{\Contr_0,\Contr_1}) \cup \Contr_2), \dots}$.
  The first equivalence is by definition of realizability
  (\cref{def:realizability}), while the second equivalence follows by the
  fact that realizability games are zero
  sum~\cite{pnueli1989synthesis,ehlers2013symmetric}. The last line
  corresponds to the nonexistence of a winning strategy of Environment for
  $\lnot\phi$.

  Crucially, the (non)existence of a winning strategy of Environment for
  any formula $\psi\in\cosafetyltl$ can be checked by using
 classical
 realizability (\cref{def:realizability}). It suffices
  to:
  \begin{enumerate*}[label=(\roman*)]
    \item swap the controllable and uncontrollable variables of $\phi$;
    \item codify the fact that Environment has to move as the second
      player.
  \end{enumerate*}
  We now show how to solve the second point. For any $\psi\in\cosafetyltl$
  with controllable variables $\Cset$ and uncontrollable variables $\Uset$,
  we define $\mealy(\psi)$ as the formula obtained from $\psi$ by replacing
  each variable $u\in\Uset$ with the formula $\ltl{X}(u)$.

  From now on, given a strategy $t : (2^\Uset)^* \to 2^\Cset$ and any
  $\Uncontr = \seq{\Uncontr_0,\Uncontr_1,\dots} \in
  (2^\Uset)^\omega$, we denote as $\res(t,\Uncontr)$ the sequence
  $\seq{(t(\epsilon) \cup \Uncontr_0), (t(\seq{\Uncontr_0}) \cup
  \Uncontr_1), (t(\seq{\Uncontr_0,\Uncontr_1}) \cup \Uncontr_2), \dots}$,
  that is the sequence of rounds in which Controller is the first to play.
  If instead $t : (2^\Uset)^+ \to 2^\Cset$, then we use the definition in
  the preliminaries (that is, Environment is the first to play):
  $\res(t,\Uncontr)$ is the sequence $\seq{(\Uncontr_0 \cup
  t(\Uncontr_0)), (\Uncontr_1 \cup t(\seq{\Uncontr_0,\Uncontr_1})),
  (\Uncontr_2 \cup t(\seq{\Uncontr_0,\Uncontr_1,\Uncontr_2}) , \dots}$.

  In the following, we prove that:
  \begin{align}
  \begin{split}
  \label{eq:mealy}
    &\exists t : (2^\Uset)^* \to 2^\Cset \suchdot \forall
    \Uncontr \in (2^\Uset)^\omega \suchdot
    \res(t,\Uncontr) \models \psi \\
    &\qquad\qquad\Iff \\
    &\exists t' : (2^\Uset)^+ \to 2^\Cset \suchdot \forall
    \Uncontr \in (2^\Uset)^\omega \suchdot
    \res(t',\Uncontr) \models \mealy(\psi)
  \end{split}
  \end{align}
  We first prove the left-to-right direction.  Suppose that $\exists
  t : (2^\Uset)^* \to 2^\Cset \suchdot \forall \Uncontr \in
  (2^\Uset)^\omega \suchdot \res(t,\Uncontr) \models \psi$. We define
  the strategy $t' : (2^\Uset)^+ \to 2^\Cset$ as follows: for any
  $\seq{\Uncontr_0,\dots,\Uncontr_n}\in(2^\Uset)^+$
  \begin{align*}
    t'(\seq{\Uncontr_0,\dots,\Uncontr_n}) =
    t(\seq{\Uncontr_0,\dots,\Uncontr_{n-1}})
  \end{align*}
  where $\seq{\Uncontr_0,\dots,\Uncontr_{n-1}}$ is the empty word
  $\epsilon$ if $n-1<0$.

  We prove the left-to-right direction by proving a stronger result. In
  order to that we first need an additional definition: for
  any $\Uncontr = \seq{\Uncontr_0,\Uncontr_1,\dots} \in
  (2^\Uset)^\omega$, we define $\mealy(\Uncontr)$ the sequence
  $\seq{\Uncontr',\Uncontr_0,\Uncontr_1,\dots}$, where $\Uncontr'$ is an
  arbitrary member of $2^\Uset$.
  We now prove the following stronger result: for any $\Uncontr
  \in (2^\Uset)^\omega$ and for any $i\in\N$, if $\res(t,\Uncontr),i
  \models \psi$, then $t'(\mealy(\Uncontr)),i \models
  \mealy(\psi)$.  
  We remark that, since by definition any uncontrollable variable in
  $\mealy(\psi)$ is prefixed by a $\ltl{X}$ operator, proving that
  $t'(\mealy(\Uncontr)) \models \psi$ is equivalent to prove
  that $\res(t',\Uncontr) \models \psi$.
  We proceed by induction on the structure of $\psi$.
  \begin{itemize}
    \item If $\psi = u$ with $u\in\Uset$, then by hypothesis we have
      that $\res(t,\Uncontr),i \models u$, that is $u \in
      \res(t,\Uncontr)_i$ and, since $u\in\Uset$, we also know that
      $u\in\Uncontr_i$.  By construction of $\mealy(U)$, this
      means that $u\in \mealy(\Uncontr)_{i+1}$.  By definition
      of $t'$, $t'(\mealy(\Uncontr)) = t(\mealy(\Uncontr))$
      and thus (in particular) $t'(\mealy(\Uncontr))_{i+1}
      = t(\mealy(\Uncontr))_{i+1}$. Therefore, $u\in
      t'(\mealy(\Uncontr))_{i+1}$ and $t'(\mealy(\Uncontr)),i
      \models \ltl{X}(u)$, that is $t'(\mealy(\Uncontr)),i \models
      \mealy(\psi)$.
    \item The proof for $\psi = \lnot u$ (with $u\in\Uset$) is
      identical to the previous point.
    \item If $\psi = c$ with $c\in\Cset$, then by hypothesis we have
      that $\res(t,\Uncontr),i\models c$, that is $c\in
      \res(t,\Uncontr)_i$. Since $c\in\Cset$, we know that $c\in
      t(\seq{\Uncontr_0,\dots,\Uncontr_i})$. By definition of $t'$, this
      means that $c\in t'(\seq{\Uncontr_0,\dots,\Uncontr_{i+1}})$ and, by
      definition of $\res(t',\Uncontr)$, this means that $c\in
      \res(t',\Uncontr)_i$. Therefore, $\res(t',\Uncontr),i
      \models c$, that is $\res(t',\Uncontr),i \models \mealy(c)$.
    \item The proof for $\psi = \lnot c$ (with $c\in\Cset$) is
      identical to the previous point.
    \item If $\psi = \phi_1\land\phi_2$, then by hypothesis
      $\res(t,\Uncontr),i \models \psi$, that is
      $\res(t,\Uncontr),i \models \psi_1$ and
      $\res(t,\Uncontr),i \models \psi_2$. By inductive hypothesis,
      $\res(t',\mealy(\Uncontr)),i \models \mealy(\psi_1)$ and
      $\res(t',\mealy(\Uncontr)),i \models \mealy(\psi_2)$, that is
      $\res(t',\mealy(\Uncontr)),i \models \mealy(\psi)$.
    \item The cases for $\psi = \phi_1\lor\phi_2$,
      $\psi = \ltl{X\psi_1}$, $\psi = \ltl{Y\psi_1}$,
      $\psi = \ltl{\psi_1 U \psi_2}$ and $\psi = \ltl{\psi_1
      S \psi_2}$ can be simply be proved by induction as in the previous
      point.
  \end{itemize}
  The right-to-left direction, that is if $t(\mealy(\Uncontr)),i
  \models \mealy(\psi)$ then $\res(t',\Uncontr),i \models \psi$ (for
  all $i>0$), can be proved similarly. This concludes the proof for
  \cref{eq:mealy}.

  Now, we go back to \cref{eq:proof:dual} and we show that the
  (non)existence of a strategy of Environment for $\lnot\phi$ can be solved
  by classical realizability (as defined in \cref{def:realizability}). Let
  $\phi$ be the formula of \cosafetyltl as defined in \cref{eq:proof:dual}
  and let $\Cset$ and $\Uset$ be the set of controllable and uncontrollable
  variables, respectively.
  We define $\dual(\phi)$ as the formula $\mealy(\not\phi)$ whose set
  $\Cset'$ of controllable variables is $\Uset$ and whose set of
  uncontrollable variables is $\Cset$. By the properties of $\mealy(\cdot)$
  that we showed before, it holds that:
  \begin{align*}
    &\lnot\exists s' : (2^\Cset)^* \to 2^\Uset \suchdot \forall
    \Contr \in (2^\Cset)^\omega \suchdot \res(s',\Contr)
    \models \lnot\phi \\
      \Iff \
    &\lnot (\exists t : (2^{\Uset'})^+ \to 2^{\Cset'} \suchdot \forall
    \Uncontr\in (2^\Uset)^\omega \suchdot \res(t,\Uncontr)
    \models \dual(\phi)) \\
      \Iff \
    &\dual(\phi) \mbox{ is not realizable }
  \end{align*}
  By \cref{eq:proof:dual}, it follows that $\phi$ is realizable iff
  $\dual(\phi)$ is not realizable. Crucially, if $\phi$ is a formula of
  \cosafetyltl, then $\dual(\phi)$ is a formula in \safetyltl of size
  polynomial in the size of $\phi$. This allows to have a reduction from
  realizability of \cosafetyltl over infinite traces to realizability of
  \safetyltl over infinite traces, and \viceversa. In particular, for any
  $\phi\in\cosafetyltl$, it holds that $\phi$ is realizable iff
  $\dual(\phi)$ is not realizable. Therefore, the realizability (over
  infinite traces) of \cosafetyltl is $\mathsf{C}$-complete iff
  realizability (over infinite traces) \safetyltl is
  $\mathsf{coC}$-complete.
  Since \LTLxf is a syntactic fragment of \cosafetyltl, all these results
  holds for \LTLxf as well.

  We now consider the case of \Falpha. Let $\ltl{F}(\phi)$ be a formula of
  \Falpha where $\phi\in\LTLFP$. Consider $\ltl{G}(\dual(\phi))$. By the
  properties of $\dual(\cdot)$, it holds that $\ltl{F(\phi)}$ is realizable
  iff $\ltl{G}(\dual(\phi))$ is not realizable. However, since
  $\dual(\phi)$ introduces additional $\ltl{X}$ operators, $\dual(\phi)$ is
  \emph{not} a \Galpha formula. We perform three equivalence-preserving
  translation for translating $\ltl{G}(\dual(\phi))$ into a formula in
  \Galpha.

  Let $\beta$ be any formula of \LTLP whose only temporal operators are
  past or the $\ltl{X}$ operator (like $\dual(\phi)$), and let $i$ be the
  maximum number of nested $\ltl{X}$ operators in $\beta$.  By using the
  \emph{pastification
  method}~\cite{maler2007synthesizing,cimatti2021extended}, without the
  addition of auxiliary variables, one can transform $\beta$ into the form
  $\ltl{X^i}(\beta')$ such that $\beta'\in\LTLFP$ and the size of $\beta'$
  is polynomial in the size of $\phi$.
  Since the maximum number of nested $\ltl{X}$ in $\dual(\phi)$ is $1$, by
  applying pastification on $\dual(\phi)$ we obtain a formula
  $\ltl{X}(\alpha')$ such that $\alpha'\in\LTLFP$ and $\ltl{G}(\dual(\phi))
  \equiv_I \ltl{G X}(\alpha')$. Now we have that:
  \begin{align*}
    \ltl{G X}(\alpha')
      \ \equiv_I \
    \ltl{X G}(\alpha')
      \ \equiv_I \
    \ltl{G(\ltl{wY\false \lor \alpha'})}
  \end{align*}
  Let $\phi'\coloneqq \ltl{G(\ltl{wY\false \lor \alpha'})}$. We have that
  $\phi$ is realizable iff $\phi'$ is not realizable. Crucially, $\phi'$ is
  a formula in \Galpha of size polynomial in the size of $\phi$. This
  witness the existence of a reduction from \Falpha realizability (over
  infinite traces) to \Galpha realizability (over infinite traces), and
  \viceversa. Therefore, the realizability (over infinite traces) of
  \Falpha is $\mathsf{C}$-complete iff realizability (over infinite traces)
  \Galpha is $\mathsf{C}$-complete.
\end{proof}

\thmrealcontraction*
\begin{proof}
  The right-to-left direction is straightforward.  
  For the opposite direction, suppose that $\phi$ is realizable. By
  \cref{def:realizability}, there exists a strategy $s : (2^\Uset)^+ \to
  (2^\Cset)$ such that, for any $\Uncontr\in(2^\Uset)^\omega$, there exists a
  $k\in\N$ for which $\res(s,\Uncontr)_{[0,k]} \models \phi$, where
  $\res(s,\Uncontr)_{[0,k]}$ is the prefix of $\res(s,\Uncontr)$ from $0$ to
  $k$.
  Since
  $\phi\in\mathbb{L}$,
  with
  $\mathbb{L} \in \{ \LTLwxg, \safetyltl, \Galpha\}$,
  by
  \cref{thm:smp:safety:fin:logic}, it holds that $\res(s,\Uncontr)_0 \models
  \phi$ for any $\Uncontr\in(2^\Uset)^\omega$. This proves the left-to-right
  direction.
\end{proof}









\cleardoublepage

\fontsize{11}{14}\selectfont

\onecolumn
\section*{\LARGE Erratum}

\vspace{0.75cm}

\setcounter{section}{0}


\section*{Introduction}
\label{sec:introduction}


This erratum addresses a mistake shared by the proofs of Lemma 2 and Lemma 8 from
the original version of the paper above,
respectively on the $\mathsf{PSPACE}$-completeness of $\cosafetyltl$ satisfiability over infinite traces, and on the $\mathsf{2EXPTIME}$-completeness of $\cosafetyltl$ realizability over infinite traces.
As we will argue in the following, the statements of both lemmas are still correct. However, their original proofs rely on reductions that cannot be carried out in polynomial time, hence leading to incorrect hardness proofs for the corresponding complexity classes.
We thank Noel Arteche for pointing out this mistake (see also~\cite{arteche2024towardsexact} for further discussion).

Regarding Lemma 2, we observe that the lower bound follows from the results obtained in~\cite{markey04pastfree}, in particular Corollary 15, showing the $\mathsf{PSPACE}$-completeness of
formula satisfiability over infinite traces of
$\LTL$ with only the until operator (and formulas in negation normal forms).
From this, we immediately obtain that $\cosafetyltl$ formula satisfiability over infinite traces is $\mathsf{PSPACE}$-hard, as required.
%
For Lemma 8, the proof adjustments required more radical interventions, leading to a novel $\mathsf{2EXPTIME}$-hardness proof that we detail in the following.

\section*{Preliminaries}
\label{sec:preliminaries}

We define the realizability problem 
for temporal logic formulas
as a two-player game between \emph{Controller}, whose aim is to satisfy the
formula, and \emph{Environment}, who tries to violate it. In this setting, the
notion of strategy plays a crucial role.

\begin{definition}[Strategy]
\label{def:strategy}
  Let $\Sigma=\Cset\cup\Uset$ be a set of variables partitioned into
  \emph{controllable}, $\Cset$, and \emph{uncontrollable}, $\Uset$, ones.
  A \emph{winning strategy} (or simply \emph{strategy}) for \emph{Controller} is a function $s:(2^\Uset)^+ \to
  2^\Cset$ that, for any finite sequence
  $\Uncontr=\seq{\Uncontr_0,\ldots,\Uncontr_n}$ of choices by
  \emph{Environment}, determines the choice $\Contr_n=s(\Uncontr)$ of
  \emph{Controller}.
\end{definition}

Let $s:(2^\Uset)^+ \to 2^\Cset$ be a strategy and let $\Uncontr=\seq{\Uncontr_0,\Uncontr_1,\ldots}$ $\in(2^\Uset)^\omega$ be an infinite sequence of
choices by \emph{Environment}. We denote
by $\res(s,\Uncontr)= \seq{\Uncontr_0\cup s(\seq{\Uncontr_0}), \Uncontr_1
\cup s(\seq{\Uncontr_0, \Uncontr_1}), \ldots}$ the state sequence resulting
from reacting to $\Uncontr$ according to $s$.  The realizability problem
can be defined as follows.

\begin{definition}[Realizability]
\label{def:realizability}
  Let $\phi$ be
  an $\LTLP$
  formula over the alphabet $\Sigma = \Cset \cup
  \Uset$, with $\Cset\cap\Uset=\emptyset$. We say that $\phi$ is
  \emph{realizable over infinite} (resp., \emph{finite}) \emph{traces} if
  and only if there exists a strategy $s : (2^\Uset)^{+} \to 2^\Cset$ such
  that, for any infinite sequence $\Uncontr=\seq{\Uncontr_0, \Uncontr_1,
  \dots}$ in $(2^\Uset)^\omega$, it holds that $\res(s,\Uncontr) \models
  \phi$ (resp., there exists $k\in\N$ such that the prefix of
  $\res(s,\Uncontr)$ from $0$ to $k$ is a model of $\phi$).
\end{definition}

Given a set of formulas $\mathbb{L}$,
the
\emph{realizability problem}
for
$\mathbb{L}$
is the problem of establishing, given a formula $\phi\in\mathbb{L}$, whether
$\phi$ is realizable.
We recall some results in the literature on the complexity
of the realizability problem of (fragments of) \LTL and \LTLP over infinite and
finite traces.

\begin{proposition}[\cite{pnueli1989synthesis,rosner1992modular,DeGiacomoV15}]
  Realizability  for \LTL and
  \LTLP over infinite and over finite traces is \EXPTIME[2]-complete.
\end{proposition}

\section*{Hardness proof}
\label{sec:proof}

This section proves that \cosafetyltl realizability is \EXPTIME[2]-complete.
Since the upper bound comes from \LTLf~\cite{DeGiacomoV15}, we focus on the
lower bound.

\subsection{Tiling games}

We prove the lower bound by a reduction from \emph{exponential corridor tiling
games}, described by Chlebus~\cite{chlebus1986domino}. We now recap tiling
problems and the specific game variant we use.
\begin{definition}[Tilings]
    \label{def:tiling}
    A \emph{tiling structure} is a tuple $\T = \tuple{T,H,V,t_0,t_F}$ where:
    \begin{enumerate}
        \item $T$ is a finite set of elements called \emph{tiles};
        \item $H:T\times T$ and $V:T\times T$ are the \emph{horizontal} and     
            \emph{vertical} adjacency relations, respectively;
        \item $t_0\in T$ and $t_F\in T$ are the \emph{initial} and \emph{final}
            tiles, respectively.
    \end{enumerate}
    Given $n,m\in\N_+$, an $n\times m$-tiling is a map
    $f:\set{0,\ldots,n-1}\times\set{0,\ldots,m-1}\to T$, which tessellates an
    $n\times m$ grid with tiles in such a way that:
    \begin{enumerate}
        \item $f(0,0)=t_0$;
        \item $f(n-1,m-1)=t_F$;
        \item for all $0\le i < n-1$ and $0\le j < m$, we have
            $(f(i,j),f(i+1,j))\in H$;
        \item for all $0\le i < n$ and $0\le j < m-1$, we have
            $(f(i,j),f(i,j+1))\in V$;
    \end{enumerate}
\end{definition}

\begin{definition}[Exponential corridor tiling problem]
    Given a tiling structure $\T$ and a height $m>0$, \emph{encoded in
    binary}, the \emph{exponential corridor tiling problem} asks to find where an
    $n\times m$-tiling for $\T$ exists, for some $n>0$.
\end{definition}

Tiling problems~\cite{van1997convenience} are well-known convenient tools for
reductions because of their strict connection with Turing machines. Each column
of a tiling can be seen as the content of the tape at a given execution step,
and the $H$ and $V$ relations can be used to encode the machine's transition
relation. Then, fixing the height of a tiling corresponds to bounding the
\emph{space} used by the machine's execution, while fixing the width corresponds
to bounding the \emph{time}. For this reason, tiling problems can easily capture
many different \emph{nondeterministic} complexity classes. Note that the length
$m$ is encoded in binary, so we get the following.
\begin{proposition}[Complexity of tilings~\cite{van1997convenience}]
    \label{prop:tilings}
    The \emph{exponential corridor tiling problem} is $\EXPSPACE$-complete.
\end{proposition}

In tiling \emph{games}, the problem is lifted to the setting of a two-player
game.

\begin{definition}[Tiling games~\cite{chlebus1986domino}]
    A \emph{tiling game} is a two-player game between \emph{Constructor} and
    \emph{Saboteur} that works as follows:
    \begin{enumerate}
        \item players are given a tiling structure $\T$ and a height
        $m>0$;
        \item players play in strictly alternating turns;
        \item \emph{Constructor} plays first;
        \item at each turn, the current player places a tile $t\in T$ for the
            current position $(i,j)$ and the game continues at next turn to
            position $(i,j+1)$, if $j<m$, or $(i+1,0)$, otherwise (\ie the tiling
            is filled column-by-column).
    \end{enumerate}
\end{definition}
\begin{definition}[Exponential corridor tiling game]
    In the \emph{exponential corridor tiling game}, the height $m>0$ is
    given, \emph{encoded in binary}, and \emph{Constructor} has the objective of
    building an $n\times m$-tiling for $\T$ for some $n>0$.
\end{definition}

In this game, \emph{Constructor} \emph{wins} the game if it has a strategy
to choose the next tile at each turn to find the required tiling in
a finite amount of steps. Formally, a strategy of \emph{Constructor} is
a function $z : (T)^* \to T$ such that,
for all \emph{Saboteur}'s choices
$S \in (T)^\omega$, it holds that the play $\seq{z(\epsilon), S_0, z(S_0),
S_1, z(S_0,S_1), \dots}$ forms a correct tiling.
for all \emph{Saboteur}'s choices
$S = \langle S_{0}, S_{1}, \ldots \rangle \in (T)^\omega$, it holds that the play $\seq{z(\epsilon), S_0, z(\langle S_0 \rangle),
S_1, z(\langle S_0,S_1 \rangle), \dots}$ forms a correct tiling (in the following, to improve readability, we often omit angle brackets).

\begin{proposition}[Complexity of tiling games]
    \label{prop:tiling:games}
    Deciding whether
    \emph{Constructor} wins an \emph{exponential corridor tiling game}
    is \AEXPSPACE-complete, \ie \EXPTIME[2]-complete.
\end{proposition}

We can see the alternating bound of \cref{prop:tiling:games} is essentially the
same as \cref{prop:tilings}. This is because the underlying reduction from
Turing machines is the same, only lifted to an alternating setting because of
the game dynamics. As a result, \cref{prop:tiling:games} gives us convenient
ways to prove $\EXPTIME[2]$ lower bounds.

\subsection{The general idea}

We will show a reduction from the exponential corridor tiling game to
\cosafetyltl realizability, by building a formula $\phi_{\T,m}$ that is
realizable iff \emph{Constructor} wins the game of height $m$ on $\T$. Note
that a naive approach would be to look for a formula that is
\emph{satisfiable} if and only if $\T$ has a corridor tiling, and then lift
the encoding to a game setting. This would work by encoding tilings as
words, column-by-column, and enforcing vertical and horizontal adjacency
relations through suitable temporal formulas.  However, such an encoding
cannot possibly work because $\cosafetyltl$ satisfiability is only
\PSPACE-complete~\cite{sistla1985complexity}, compared to the exponential
corridor tiling problem which is \EXPSPACE-complete. What breaks down is that,
since $m$ is encoded in binary, columns are exponentially long, and
therefore enforcing of the horizontal adjacency relation is impossible with
a polynomially-sized formula. The naive approach works instead for tiling
games where $m$ is given in \emph{unary}, where the corridor tiling problem
is indeed \PSPACE-complete~\cite{van1997convenience}.

The same problem appears in any similar proof for full \LTL, so we get
inspiration from the technique employed by Pnueli and Rosner~\cite{PnueliR89},
who in turn got inspired by the \EXPTIME[2] lower bound of satisfiability of the
\CTL* branching-time logic by Vardi and Stockmeyer~\cite{VardiS85}.

In our reduction, a \emph{counter} of $\ceil{\log_2(m)}$ bits is used to
keep track of the current row in the tiling, which is represented linearly
column-by-column by the word. In this setting, the vertical adjacency
relation is easy to enforce with a simple \emph{tomorrow} operator.
Instead, the horizontal one is tricky, because we cannot refer to the next
$2^m$ time point with a polynomial-sized formula. To enforce the horizontal
relation, the technique inspired by Vardi and Stockmeyer~\cite{VardiS85}
predicates on \emph{every subinterval} of the trace up to the end of the
tiling. Only in subintervals when the row counter happen to have the same
value at the start and at the end of the interval, we ask for the tiles at
those specific points to be compatible. Crucially, it is the
\emph{Environment} player who chooses the last point of the subinterval.
The resulting formula will accept models encoding valid tilings, but also
many other invalid ones. However, since \emph{Controller} cannot know
\apriori if and when \emph{Environment} will choose to end the subinterval,
the only safe strategy is to always build a correct tiling in the first
place.

\subsection{The reduction}

Let us now give more details. Let $\T=\tuple{T,H,V,t_0,t_F}$ and $m>0$. We build
the encoding formula $\phi_{\T,m}$ upon the following set of propositions
$\AP=\Contr\cup\Uncontr$:
\begin{align*}
    \Contr & =
        \underbrace{\set{b_i^c\mid 0\le i < \ceil{\log_2(\abs{T})}}}_{\text{tiles}}\cup 
        \underbrace{\set{c_i \mid 0\le i < \ceil{log_2(m)}}}_{\text{counters bits}} \cup 
        \underbrace{\set{u}}_{\text{turn \rlap{marker}}}\\
    \Uncontr & =
        \underbrace{\set{b_i^u\mid 0\le i < \ceil{\log_2(\abs{T})}}}_{\text{tiles}}\cup
        \underbrace{\set{\mathit{end}}}_{\text{end of \rlap{the interval}}}
\end{align*}

Counter bits $c_i$ track the current row. We can test the value of the
counter by testing the bits individually, denoting it as $c=k$ for brevity.
This can be done with a Boolean formula of size polynomial in $\log_2(c)
+ \log_2(k)$. We use the notation $c\coloneqq c+1$ to denote a Boolean
formula encoding the fact that at the next state the counter increments by
$1$ \emph{modulo $m$}, \ie it wraps from $m-1$ to $0$. Such
a polynomial-size formula can be built in standard ways.

The current tile is represented by $\log_2(\abs{T})$ bits, in an arbitrary
Boolean encoding, and we use symbols $t^u$ and $t^c$ for some $t\in T$ to say
that the bits $b_i^u$ and $b_i^c$, respectively, correspond to $t$. 

As per the definition of \cosafetyltl realizability, at each turn of the game
playing $\phi_{\T,m}$, each player chooses how to play their propositions, which
include $t_i^c$ for \emph{Controller} and $t_i^u$ for \emph{Environment},
for all $0\le i < \ceil{\log_2(\abs{T})}$. 
Each turn in the tiling game is a different temporal step in the models of
the formula. Whose player is the current turn is kept track by the $u$
proposition which toggles at each step and is true when its
\emph{Environment}'s turn to play. 
Moreover, we will construct the formula $\phi_{\T,m}$ in such a way that
the variables in $\set{b_i^c\mid 0\le i < \ceil{\log_2(\abs{T})}}$ (resp.,
the variables in $\set{b_i^u\mid 0\le i < \ceil{\log_2(\abs{T})}}$) are not
constrained in any way (\ie they are \emph{don't care} variables) if the
current turn belongs to Environment (resp., belongs to Controller).
This has also the following advantages: 
\begin{enumerate}[label=(\roman*)]
  \item who plays first in the single step is irrelevant;
  \item the mismatch in the fact that \emph{Constructor} plays first in the
    tiling game but \emph{Environment} plays first in the realizability
    game does not affect the encoding;
  \item the formula $\phi_{\T,m}$ does not have to force the fact that,
    when the turn belongs to Controller (resp., to Environment), exactly
    one variable among $\set{b_i^c\mid 0\le i < \ceil{\log_2(\abs{T})}}$
    (resp., $\set{b_i^u\mid 0\le i < \ceil{\log_2(\abs{T})}}$) is played.
\end{enumerate}

The formula $\phi_{\T,m}$ is defined as follows:
\begin{equation*}
    \phi_{\T,m}\coloneqq 
    \ltl{\phi_{\mathit{init}}} \land
        \ltl{(\phi_c \land \phi_h \land \phi_v) U \phi_{\mathit{goal}}}
\end{equation*}
where:
\begin{enumerate}
    \item
      $\phi_{\mathit{init}} \coloneqq t^c_0 \land c=0 \land \neg u$ states that $t_0$ (the initial tile
      in~\cref{def:tiling}) is the first tile to be played, that the counter
      starts at zero, and that is \emph{Controller}'s turn to play;
    \item $\phi_{\mathit{goal}}\coloneqq (c = m-1 \land t_F^c)$
        requires the existence of a column with final tile at the top;
    \item $\phi_c\coloneqq (c\coloneqq c+1) \land (u\iff\ltl{X\neg u})$
      ensures the correct behavior of the counter bits and of the turn
      marker bit $u$;
    \item $\phi_v$ enforces the vertical relation and is defined as
        $\phi_v\coloneqq \phi_v^u \land \phi_v^c$ where:
        \begin{align*}
            \phi_v^u & \coloneqq (u \land c\ne m-1) \implies 
                \smashoperator{\bigvee_{(t_1,t_2)\in V}} 
                    (t_1^u \land \ltl{X t_2^c}) & 
            \phi_v^c & \coloneqq (\neg u \land c\ne m-1) \implies 
                \smashoperator{\bigvee_{(t_1,t_2)\in V}} 
                    (t_1^c \land \ltl{X t_2^u})
        \end{align*}
    \item\label{cond:intervals} 
        $\phi_h$ enforces the horizontal relation only between the
        endpoints of the subinterval whose starting point is the current
        time point on which $\phi_h$ is interpreted, and the endpoint is
        either the position (if any) preceding one where
        \emph{Environment} chooses to play $\mathit{end}$ or the position
        in which $\phi_{\mathit{goal}}$ holds. This is expressed by
        a disjunction of the following. Either:
        \begin{enumerate}
            \item\label{cond:intervals:inside}
              the current interval includes a single time point:
                \begin{equation*}
                  \phi_{\mathit{goal}} \lor \ltl{X}(\mathit{end})
                \end{equation*}
            \item\label{cond:intervals:multiple} 
                the current interval spans at least two runs of the row
                counter:
                \begin{equation*}
                    \ltl{F (c = 0 \land X F(c = 0 \land 
                      X F(\phi_{\mathit{goal}} \lor \ltl{X}(\mathit{end}))))}
                \end{equation*}
            \item\label{cond:intervals:different}
                the value of the counter at the beginning and at the end of
                the interval is different (note that this includes also the
                case in which the interval spans only a single run of the
                counter):
                \begin{equation*}
                    \ltl{
                        \neg\bigwedge\nolimits_{i=0}^{\ceil{\log_2(m)}}
                        (c_i \iff F(c_i \land (\phi_{\mathit{goal}} \lor X(\mathit{end}))))
                    }
                \end{equation*}
            \item\label{cond:intervals:matching}
                or, the tiles at the start and end of the interval match
                (accounting for the right turns):
                \begin{equation*}
                    \bigvee_{(t_1,t_2)\in H} 
                    \left\{
                    \begin{aligned}
                        u\land t_1^u &{}\land 
                          \ltl{F(u\land t_2^u\land 
                            (\phi_{\mathit{goal}} \lor X \mathit{end}))} \lor {} \\
                        u\land t_1^u &{}\land 
                          \ltl{F(\neg u\land t_2^c\land 
                            (\phi_{\mathit{goal}} \lor X \mathit{end}))} \lor {} \\
                        \neg u\land t_1^c &{}\land 
                          \ltl{F(u\land t_2^u\land 
                            (\phi_{\mathit{goal}} \lor X \mathit{end}))} \lor {} \\
                        \neg u\land t_1^c &{}\land 
                          \ltl{F(\neg u\land t_2^c\land 
                            (\phi_{\mathit{goal}} \lor X \mathit{end}))}
                    \end{aligned}
                    \right.
                \end{equation*}
        \end{enumerate}
\end{enumerate}

To intuitively understand the encoding of condition \ref{cond:intervals},
consider it as stating that the negations of conditions
\ref{cond:intervals:inside}, \ref{cond:intervals:multiple}, and
\ref{cond:intervals:different} together imply condition
\ref{cond:intervals:matching}. 
Moreover, note that, in each run of the game, even if \emph{Environment}
plays $\mathit{end}$ somewhere, there is no obligation for
\emph{Controller} to build a correct tiling \emph{in that particular run}.
However, since the existence and the placement of $\mathit{end}$ is not
known, the only strategy that ensures to be ready to fulfil condition~\ref{cond:intervals:matching})
above \emph{at any time} is to \emph{always} build a correct tiling. It is
easy to see that $\phi_{\T,m}$ can be produced in polynomial time,
therefore we are left to confirm that $\phi_{\T,m}$ is realizable if and
only if \emph{Constructor} wins the game, proving the following.
\begin{theorem}
    \cosafetyltl realizability is \EXPTIME[2]-complete.
\end{theorem}
\begin{proof}
    We start by proving that if \emph{Constructor} wins the tiling game,
    then \emph{Controller} has a winning
    strategy for $\phi_{\T,m}$. We
    start by noticing that, among the propositions owned by
    \emph{Controller}, the counter bits can only ever evolve in a single
    predefined way, so their choices are fixed.  What \emph{Controller}
    really has to choose are the $t_i^c$ propositions. It is
    straightforward to define the strategy $s:(2^\Uset)^+\to(2^\Cset)$
    that:
    \begin{enumerate}
        \item to choose the values of the $t_i^c$ propositions, replays the
            moves of \emph{Constructor} in the tiling game;
        \item chooses the counter bits to suitably represent a binary counter 
            modulo $m$;
        \item strictly alternates between $u$ and $\neg u$ at each step.
    \end{enumerate}
    Note that the choices of $s$ never depend on the choice of
    \emph{Environment} of where to play $\mathit{end}$, if at all. Since
    \emph{Constructor} wins the tiling game, we can check that $s$ is
    a winning strategy for $\phi_{\T,m}$, that is, $\res(s,\Uncontr)
    \models \phi_{\T,m}$ for all $\Uncontr \in (2^\Uset)^\omega$, because:
    \begin{enumerate}
        \item $t_0$ is mandatorily the first tile to be played by
            \emph{Constructor}, therefore $t_0^c$ is also the first move played
            by \emph{Controller}, and the choices for the counter bits and $u$
            are fixed, so $c=0$ and $\neg u$ are satisfied as well;
        \item the contructed tiling is a valid tiling for $\T$, therefore
          there is a column with $t_F$ at the top, which means there is
          a position in $\res(s,\Uncontr)$ where both $c=m-1$ (top
          position), and $t_F^c$ has been played by \emph{Controller}; this
          means the existential requirement of the \emph{until} is
          satisfied and we are left to ensure the universal requirement
          (\ie $\phi_c \land \phi_h \land \phi_v$) is fulfilled at all
          steps until then;
        \item the satisfaction of $\phi_c$ is guaranteed by construction of
          the strategy $s$;
        \item the satisfaction of the vertical contraint in the tiling is
            guaranteed by it being a valid tiling, and this reflects directly on
            the satisfaction of $\phi_v$, and in particular of $\phi_v^u$ or
            $\phi_v^c$ in \emph{Environment}'s and \emph{Controller}'s turns,
            respectively;
        \item let $i$ be \emph{any} step in $\res(s,\Uncontr)$. Then, let
          $k\ge i$ be the smallest among the first position where
          \emph{Environment} plays $\mathit{end}$ in the next step (if any)
          and the position where $\phi_{\mathit{goal}}$ holds. Then:
            \begin{enumerate}
                \item if $k=i$, then condition \ref{cond:intervals:inside} is
                    satisfied;
                \item if the counter resets to zero more than once between the
                    two positions, condition \ref{cond:intervals:multiple} is
                    satisfied;
                \item if the values of the counter in the two positions are
                  different, condition \ref{cond:intervals:different} is
                  satisfied;
                \item otherwise, the only remaining case is when $k>i$, the
                  counter resets exactly one time, and the values of the
                  counter at the beginning and at the end of the interval
                  are equal, hence condition \ref{cond:intervals:matching} is satisfied: this is possible only when the two positions
                  are representing a horizontally adjacent pair of tiles.
                  In this case, the horizontal adjacency relation is
                  guaranteed to hold by the tiling produced by
                  \emph{Constructor} being a valid one.
            \end{enumerate}
    \end{enumerate}

    Let us now prove the opposite direction, \ie if $\phi_{\T,m}$ is
    realizable with a strategy $s$,
    then \emph{Constructor} wins the tiling game.
    We define the strategy $z : (T)^* \to T$ for \emph{Constructor} as
    follows:
    \begin{equation*}
      z(t_0,\dots,t_{i-1}) = 
      s(\emptyset, \set{t_0^u}, \dots, \emptyset, \set{t_{i-1}^u})
        \qquad  \forall i\ge 0
    \end{equation*}
    In particular, $z$ coincides with $s$ restricted to the case in which
    \emph{Environment} never plays $\mathit{end}$.\footnote{Note that the $\emptyset$ can be replaced with any set not containing $\mathit{end}$, since it is not relevant what the Environment plays at that stage.}
    It is immediate to prove that, for any choice of tiles of
    \emph{Saboteur}, strategy $z$ ensures that:
    \begin{enumerate*}[label=(\roman*)]
      \item the tile in the bottom-left corner is $t_0$;
      \item the tile in the top-right corner is $t_F$;
      \item the vertical relation is fulfilled.
    \end{enumerate*}
    We now prove that $z$ guarantees also the fulfillment of the horizontal
    relation. Suppose by contradition that this is not the case, that is,
    there exists a play $\pi \coloneqq \seq{z(\epsilon), S_0, z(S_0), S_1,
    z(S_0,S_1), \dots}$ such that:
    \begin{enumerate}[label=(\roman*)]
      \item at some (even) position $i\ge m$ (where $m$ is the height of
        the tiling structure), \emph{Constructor} chooses tile $t$, \ie
        $z(h) = t$, where $h$ is the sequence of choices of \emph{Saboteur}
        made before $i$;
      \item at position $j=i-m$, the tile chosen was $t'$; and
      \item tiles $t$ and $t'$ are \emph{not} horizontally adjacient, \ie
        $(t,t') \not\in H$.
    \end{enumerate}
    Now consider any infinite sequence of \emph{Environment}'s choices
    $\Uncontr^{(i)} \in (2^\Uset)^\omega$ such that:
    \begin{enumerate}[label=(\roman*)]
      \item it agrees with \emph{Saboteur}'s choices in $\pi$ for all
        positions from $0$ to $i$ (in particular, this means that
        $\mathit{end}$ does not belong to any of these positions); and
      \item position $i+1$ is the only one that contains proposition
        $\mathit{end}$.
    \end{enumerate}
    By definition of $z$, in particular from the fact that, over
    $\Uncontr^{(i)}$, strategy $z$ replicates the choices of $s$, it means
    that the play $\res(s,\Uncontr^{(i)}),j \not\models \phi_h$ because:
    \begin{enumerate}[label=(\roman*)]
      \item condition \ref{cond:intervals:inside} is violated since, by
        construction of $\Uncontr^{(i)}$ and by the fact that the play
        lasted at least until $i$, the first position to fulfill
        $\phi_{\mathit{goal}} \lor \ltl{X}(\mathit{end})$ is $i+1$;
      \item condition \ref{cond:intervals:multiple} is violated, since between $j$ and
        $i$ the counter resets exactly one time;
      \item condition \ref{cond:intervals:different} is violated, since the
        counter at positions $j$ and $i$ has the same value;
      \item condition \ref{cond:intervals:matching} is violated, because
        $(t,t') \not\in H$.
    \end{enumerate}
    But this is in contradiction with $s$ being a winning strategy for
    $\phi_{\T,m}$.
    Therefore, strategy $z$ always produces tilings where tile at position
    $i$ are horizontally adjacient to tile at position $i-m$, for all $i
    \ge m$, until a correct tiling is built.
\end{proof}


\fi

\end{document}